\documentclass[aps,prb,twocolumn,nofootinbib]{revtex4-2}

\usepackage[a-1b]{pdfx}

\usepackage{graphicx}
\usepackage{dcolumn}
\usepackage{bm}
\usepackage{multirow}

\usepackage[normalem]{ulem}

\usepackage{amsmath}
\usepackage{amsthm}
\usepackage{amstext}
\usepackage{amssymb}
\usepackage{mathrsfs}
\usepackage{amsfonts}
\usepackage{amsbsy} 

\usepackage[all,matrix,cmtip]{xy}

\usepackage{csquotes}
\MakeOuterQuote{"}

\usepackage{color}

\definecolor{red}{rgb}{1,0,0}
\definecolor{blue}{rgb}{0,0,1}
\definecolor{dblue}{rgb}{0,0,0.4}
\definecolor{green}{rgb}{0,1,0}
\definecolor{dgreen}{rgb}{0,0.4,0}
\definecolor{black}{rgb}{0,0,0}
\definecolor{white}{rgb}{1,1,1}

\definecolor{brn}{rgb}{.8,.4,.0}
\definecolor{redo}{rgb}{1,.5,.0}
\definecolor{ddgrn}{rgb}{0,0.4,0}
\definecolor{dgrn}{rgb}{0,0.55,0}
\definecolor{dbl}{rgb}{0,0,0.5}
\definecolor{grey}{rgb}{0.5,0.5,0.5}

\newcommand{\red}{\color{red}}

\usepackage[bbgreekl]{mathbbol}

\newcommand{\Z}{\mathbb{Z}}

\newcommand{\N}{\mathbb{N}}

\newcommand{\onebb}{\mathbb{1}}

\newcommand{\ii}{\hspace{1pt}\mathrm{i}\hspace{1pt}}
\newcommand{\ee}{\hspace{1pt}\mathrm{e}}

\newcommand{\<}{\langle} 
\renewcommand{\>}{\rangle} 

\newcommand{\Rf}[1]{Ref.~\onlinecite{#1}}

\newcommand{\ie}{{\it i.e.~}}

\newcommand{\bpm}{\begin{pmatrix}}
\newcommand{\epm}{\end{pmatrix}}
\newcommand{\bmm}{\begin{matrix}}
\newcommand{\emm}{\end{matrix}}

\usepackage{euscript}

\newcommand{\al}{\alpha} 
\newcommand{\bt}{\beta} 
\newcommand{\del}{\delta}

\newcommand{\vphi}{\varphi} 

\newcommand{\frmbox}[1]{\begin{center}\fbox{\parbox{3.3in}{\parindent=0pt #1}}\end{center}}

\usepackage{hyperref}
\hypersetup{
    colorlinks = true,
    allcolors = blue,
}

\usepackage{tikz}

\tikzstyle{black dot}=[fill=black, draw=black, shape=circle]
\tikzstyle{black circ}=[fill=white, draw=black, shape=circle]
\tikzstyle{white circ}=[fill=white, draw=white, shape=circle]

\begin{document}

\title{Fibonacci Non-Abelian Quantum Hall States with Pair Structure\\
-- a classification of chiral 2-cluster many-body wave-functions
}

\author{Abigail Timmel}

\author{Xiao-Gang Wen}

\affiliation{Department of Physics, Massachusetts Institute of Technology,
Cambridge, Massachusetts 02139, USA
}

\begin{abstract} 

Fibonacci anyons support universal topological quantum computation. They have
been proposed to emerge in fractional quantum Hall (FQH) systems, such as the
filling-fraction $\nu=\tfrac{2}{3}$ state described by an effective $SU(3)_2$
Chern-Simons theory, and the $\nu=\tfrac{2}{5}$ state with a $3$-cluster
structure (an $n$-cluster structure means that $n$-electron bound states form
a Laughlin-type wave function of the form $(z^{(n)}_i-z^{(n)}_j)^m$). In this
work, we systematically classify FQH wave functions $P(z_1,\ldots,z_N)$ in the
$N\to\infty$ limit, through a \emph{generalized pattern of zeros} (gPOZ)
approach, where the wave function is determined by the leading order of zero
along with the accompanying leading polynomial that develops
as successive number of $z_i$'s are brought together. 
Within this approach, we obtain a $(\nu,h,c)$-classification of wave functions 
with a $2$-cluster (pair) structure. We find that all such wave functions which
are primitive under stacking (\ie are not a product of several wave functions
satisfying our conditions) can
be 1-to-1 labeled by $(h,c) = (\tfrac12,\tfrac12)$, $(\tfrac34,-\tfrac35)$,
$(1,1)$, $(\tfrac54,-\tfrac{25}7)$, $ (\tfrac32,\tfrac{7}{10})$,
$(\tfrac32,-\tfrac{21}{4})$, $(2,-\tfrac{44}{5})$,
$(\tfrac94,-\tfrac{117}{11})$, $(\tfrac52,-\tfrac{13}{14})$,
$(\tfrac{11}4,-\tfrac{187}{13})$, $(3,\tfrac45)$, $(3,-\tfrac{114}{7})$ for
$h\leq 3$, together with a $\nu$ that make $\nu^{-1}-2h$ to be a positive odd
integer.  
Among these wave functions, those with $(h,c) = (\tfrac12,\tfrac12)$,
$(\tfrac32,\tfrac{7}{10})$, $(3,\tfrac45)$ correspond to gapped FQH states
with filling fraction $\nu$ and edge-state chiral central charge $c+1$, while the
others with $c<0$ may be gapless.  We find that even a $2$-cluster structure
can give rise to Fibonacci statistics, for $(\nu,h,c) =
(\tfrac14,\tfrac32,\tfrac{7}{10})$, corresponding to the $(5,4)$ minimal
model, and for $(\nu,h,c) = (\tfrac17,3,\tfrac45)$, corresponding to the
$(6,5)$ minimal model. We further construct an ideal $4$-body Hamiltonian
whose ground state is the $(\tfrac14,\tfrac32,\tfrac{7}{10})$ state.

\end{abstract}

\maketitle

{\small
\tableofcontents \setcounter{tocdepth}{2} 
}
\section{Introduction}

The Fibonacci anyon is a non-Abelian quasiparticle capable of performing
universal topological quantum computation~\cite{FLZ0205,BS0505065}. Although
controlling, or even detecting, such excitations remains an ongoing
experimental challenge, the first step toward realizing them is to find
ground-state wave functions with the appropriate topological order. Two
well-studied cases are the $\left(\chi_2(\{z_i\})\right)^3$
state~\cite{W9102,BW9215}, proposed in 1991, where $\chi_n(\{z_i\})$ is the
many-electron wave function of $n$ filled Landau levels, and the
$\mathbb{Z}_3$ parafermion state~\cite{RR9984}. The
$\left(\chi_2(\{z_i\})\right)^3$ state is described by a low-energy effective
$SU(3)_2$ Chern-Simons theory, and the $\mathbb{Z}_3$ parafermion state is
described by the $\mathbb{Z}_3$ parafermion conformal field theory (CFT); both
support Fibonacci anyons.

The $\left(\chi_2(\{z_i\})\right)^3$ state has third order zero between every
pair of electrons, but occupies the lowest four Landau levels, making it
difficult to realize in systems where the Landau levels are well separated in
energy. However, 4-layer rhombohedral stacked graphene can realize four nearly
degenerate Landau levels \cite{MM07114333}, which provide a potential
experimental realization of the $\left(\chi_2(\{z_i\})\right)^3$
state\cite{TW230809702}.

The same topological order may also be realized in the lowest Landau level by
using four electron flavors in place of the four Landau-level
indices~\cite{BW9215}. However, in this setting the exchange interaction tends
to polarize the four-component isospin, so that the $SU(3)_2$ state must
overcome a large exchange energy.

There may be evidence that the $\mathbb{Z}_3$ parafermion state appears in the
second Landau level at filling fraction $\nu=12/5$, both
experimentally~\cite{XWc0406724} and
numerically~\cite{ZS150503050,MP150502843}. The second Landau level appears to
favor bubble states~\cite{FK9326}, in which clusters of three electrons, with
Laughlin-like correlations among themselves, sit on a Wigner-crystal-like
lattice. The emergence of the $\mathbb{Z}_3$ parafermion state can be viewed
as the melting of these bubble states.

Simpler than the $\mathbb{Z}_3$ parafermion state is the $\mathbb{Z}_2$
parafermion (Pfaffian) state~\cite{MR9162}, also proposed in 1991. It hosts
Ising anyons, which are non-Abelian but do not support universal topological
quantum computation. Reference~\cite{MR9162} pioneered the approach of using
CFT correlators to construct quantum Hall states: in this case, a free-fermion
CFT produces the Pfaffian with poles, combined with a chiral-boson CFT that
contributes a Laughlin-like wave function to cancel all the poles. This approach
generalizes to $\mathbb{Z}_n$ parafermion states~\cite{RR9984}, which match
the correlators of the $\mathbb{Z}_n$ parafermion CFTs computed
in~\cite{ZF8515}.

The correlators corresponding to $\mathbb{Z}_n$ parafermion states form the
simplest examples of so-called $n$-cluster wave functions. A FQH wave function
has an $n$-cluster structure if an $n$-electron bound state has, at leading
order, a simple Jastrow form relative to all other electrons.  More precisely,
if we replace $z_i$ with $z^{(n)}+\lambda w_i$ for $i=1,\ldots n$, then the
$z^{(n)}$ dependence of the wave function appears as an overall factor
$\prod_{i>n} (z^{(n)} - z_i)^m$ as we take $\lambda \rightarrow 0$.  This is
related to the filling fraction $\nu=n/m$ since there is an average order of
$m/n$ zeros between particles.  

$n$-cluster states have been previously studied in~\cite{WW0808,WW0809} using
the pattern-of-zeros (POZ) approach, in which the states are characterized by
a sequence of integers $S_2,S_3,\ldots$, where $S_a$ is the minimal relative
angular momentum of $a$ electrons in the FQH wave function, which satisfy
\begin{align}
\label{Sacond}
& S_a \in \N, \ \ S_0=S_1=0,\ \ S_{a+b}-S_a-S_b \geq 0 .
\nonumber\\
&S_{a+b+c} -S_{a+b}-S_{b+c}-S_{a+c} +S_a+S_b+S_c \geq 0
.
\end{align}
This is analogous to characterizing a superconducting state by the relative
angular momentum of an electron pair.  However, it is known that the POZ data
$S_a$ generally fail to fully characterize FQH states: distinct FQH wave
functions can share the same POZ.  Even POZ $S_a$ plus the unique fusion
condition \eqref{factor} fail to determine the many-body wave function.  This
shortcoming is related to structure in the subleading terms, which is
unconstrained by the $n$-cluster condition, unique fusion condition, and POZ
\cite{WW0808,WW0809}.

In this work, we generalize the POZ approach for holomorphic FQH wave
functions by retaining the leading angular momentum $S_a$ along with a
leading $a$-variable polynomial at that order, describing the portion of the
wavefunction involving only those $a$ particles.  This includes information
on subleading zeros which nevertheless recover the same angular momentum
within an $a$-particle cluster.  To describe this infinite data for 
$n$-cluster states, we develop an operator expansion to produce the 
$\ell n$-particle polynomial from the $(\ell-1)n$-particle polynomial.
This expansion connects to the operator product expansion (OPE) of a
CFT, thereby connecting the CFT-correlator approach to the POZ.  The
finite data characterizing the state beyond the POZ data will be 
the central charge, $c$.  

There have been previous works relating FQH polynomials to FQH states
via the OPE~\cite{WWH9476,WW9455}.  These approaches relied on identifying
an operator to act as the energy-momentum tensor of a CFT, from which
the OPE follows using usual techniques involving conformal invariance.
Here we do not invoke conformal invariance, but we rather construct the OPE using only
conditions motivated by a lowest Landau level wave function.
In other works, the connection of the bulk wave function
to a CFT is made via correspondence with a CFT on the edge~\cite{DR245310,R0904}.  
Our arguments suggest new explanations for why this correspondence is
natural just from properties of the bulk.

We find that, beyond the leading order of the zeros, the subleading zeros are
essential for determining FQH states more complicated than the $\mathbb{Z}_n$
parafermion states~\cite{RR9984}. We provide a classification of $2$-cluster
states beyond the $\mathbb{Z}_2$ parafermion  state, corresponding to a field
of conformal weight $1/2$, up to conformal weight $3$. Our approach relies on
building a particular representation of the generators of holomorphic
transformations, which we will assume is primitive under stacking.  
A nonprimitive realization would involve decomposing each electron coordinate
$z_i$ into a product of several coordinates with different labels, 
$z_i = z_i^a z_i^b,\ldots $, acting on each of them separately, then
fusing them at the end.  Once we determine the primitive ones, however, non-primitive
combinations can be realized by multiplying multiple primitive wave
functions together.

Remarkably, we
find $2$-cluster examples that can support Fibonacci anyons. One occurs at
$\nu=1/4$ with central charge $c=7/10+1$, corresponding to the $M_{5,4}$
Virasoro minimal model plus a chiral boson. The other is a $2$-cluster
realization of the $M_{6,5}$ minimal model, occurring at $\nu=1/7$ with
central charge $c=4/5+1$.  The same minimal model is also connected to the
$3$-cluster $\mathbb{Z}_3$ parafermion state at $\nu=3/5$ with central
charge $c=4/5+1$.

\section{Generalized pattern-of-zeros approach to fractional quantum Hall
states}

\subsection{The pattern-of-zeros approach}

Fractional quantum Hall (FQH) states~\cite{TSG8259,L8395} in the first Landau
level can be described by wave functions of the form
\begin{align}
\label{Psi}
 \Psi(z_1,\cdots,z_N) =
P(\{z_i\})\, \ee^{-\frac14 \sum_{i=1}^N |z_i|^2},
 \ \ \ i=1,\cdots,N ,
\end{align}
where $N$ is the number of electrons, $z_i=x_i+\ii y_i$ is the coordinate of
the $i^\text{th}$ electron in two-dimensional space, and $P(\{z_i\})$ is an
antisymmetric polynomial for fermionic electrons. For the ground state, the
polynomial $P(\{z_i\})$ is also translationally invariant,
\begin{align}
 P(\{z_i\}) = P(\{z_i-z\}).
\end{align}
Our aim will be to characterize these antisymmetric, translation-invariant polynomials 
of $N$ variables in the limit $N\to\infty$.

To study these polynomials of infinitely many variables systematically, a
sequence of integers $\{S_a \mid a \in \Z_{>0}\}$, called the pattern of zeros
(POZ), was introduced in
\Rf{WW0808,WW0809,BW0932,BW09060337,BW1001,LWW1024,SW171105937,P180701811,P210405777}.
(We note that the ``1D charge-density-wave''
characterization of single-component FQH states~\cite{SL0604,BKW0608,ABK0816}
is closely related to the POZ approach.) The POZ approach gives a simple, but
only partial, description of the FQH wave functions.
As we bring $a$ electrons together,
\begin{align}
 z_i = \lambda\, w_i + z^{(a)} , \qquad \lambda \to 0,\ \ i =1,\ldots, a,
\end{align}
the FQH polynomial $P(\{z_i\})$ becomes
\begin{align}
P(\{z_i\}) &= \lambda^{S_a}\, P_a(w_1,\ldots, w_a)\, \times
\nonumber\\
&\qquad
P_\text{red}(z^{(a)};z_{a+1},\ldots, z_{N}) +
\mathcal{O}(\lambda^{S_a+1}).
\end{align}
We see that $S_a$ is the leading order of the zeros as $a$ electrons are
brought together. Here $P_a(w_1,\ldots, w_a)$ is a homogeneous polynomial of
degree $S_a$, which can itself be viewed as the FQH polynomial of $a$
electrons. $P_\text{red}(z^{(a)};z_{a+1},\ldots, z_{N})$ is a reduced
polynomial depending only on the cluster center coordinate $z^{(a)}$
and particles outside the cluster. $\mathcal{O}(\lambda^{S_a+1})$ denotes subleading orders
of $\lambda$.

As stated earlier, we will consider a state to have an $n$-cluster structure
if an $n$ electron bound state produces a Jastrow factor relative to all other particles,
\begin{align}
\label{factor}
P(&\{z_i\}) = \lambda^{S_n}\, P_a(w_1,\ldots, w_n)
\left(\prod_{i>n} (z^{(n)} - z_i)^m\right)
\\
&\qquad
\times P'_{N-n}(z_{n+1},\ldots, z_{N})+
\mathcal{O}(\lambda^{S_n+1}).
\nonumber 
\end{align}
In other words, the $z^{(a)}$ dependence of $P_\text{red}$ enters only as
$\prod_{i>n}(z^{(n)}-z_i)^m$, and the rest of the wavefunction separates as
an $(N-n)$-variable polynomial.
A consequence of this is that the sequence $S_a$ is completely determined by
the finite subsequence $S_2,\ldots, S_{n+1}$ (note that $S_1 = 0$).  When
we bring the $(kn+a)$th electron, it must carry $m$ zeros with
each of the $k$ $n$-clusters already present, plus $S_a$ additional zeros from
building a new $n$-electron bound state.

Because the FQH polynomial is holomorphic, $S_a$ is also the minimal relative
angular momentum among the $a$ electrons. The POZ characterization of a FQH
state is therefore analogous to the characterization of a superconductor by
the angular momentum of its electron pairs.

The POZ $S_a$ is also the total angular momentum of the $a$-electron FQH
state. From $S_a$ we obtain
\begin{align}
l_a = S_a - S_{a-1},
\end{align}
which has the physical meaning that the $a^{\text{th}}$ electron occupies the
orbital with angular momentum $l_a$. The corresponding occupation numbers are
\begin{align}
n_l = \sum_{a=1}^\infty \del_{l,l_a},
\end{align}
so that $n_l$ counts the number of electrons in the orbital with angular
momentum $l$.

The occupation pattern $n_l$ reveals the cluster structure of the FQH
polynomial. For example, the $\mathbb{Z}_3$ parafermion state has the
occupation pattern
\begin{align}
\begin{array}{|c||c|c|c|c|c||c|c|c|c|c||c|}
\hline
l   & 0 & 1 & 2 & 3 & 4 & 5 & 6 & 7 & 8 & 9 & \cdots\\
\hline
n_l & 1 & 1 & 1 & 0 & 0 & 1 & 1 & 1 & 0 & 0 & \cdots\\
\hline
\end{array}
\end{align}
Here $n_l$ is periodic, with three electrons in every five orbitals. The three
electrons per repeating block indicate a $3$-cluster structure, while the
period-five repetition means there is, on average, a $5/3$-order zero between
each pair of electrons, implying a filling fraction $\nu=3/5$.  

The filled Landau level has the occupation pattern
\begin{align}
\begin{array}{|c||c|c|c|c|c|c|c|c|c|c|c|}
\hline
l   & 0 & 1 & 2 & 3 & 4 & 5 & 6 & 7 & 8 & 9 & \cdots\\
\hline
n_l & 1 & 1 & 1 & 1 & 1 & 1 & 1 & 1 & 1 & 1 & \cdots\\
\hline
\end{array}
\end{align}
Here the repeating unit is one electron per orbital. The wave function has a
$1$-cluster structure with a $1$-order zero between each pair of electrons,
implying a filling fraction $\nu=1$.

\subsection{The generalized pattern-of-zeros approach}

Since the POZ $S_a$ fails to fully characterize the FQH polynomial, we wish to
supplement $S_a$ with additional data in the hope of obtaining a complete
characterization.  To this end, we now examine the polynomial {\it within} the
$n$-electron bound state.  

For a $n$-cluster wavefunction, again we have
\begin{align}
P(z_{1},&\ldots,  z_{N}) = \lambda^{S_n}\, P_n(w_1,\ldots, w_n)\, 
\left(\prod_{i>n} (z^{(n)}-z_i)^m\right)
\\
&
\times P'_{N-n}(z_{n+1},\ldots, z_{N})
+
\mathcal{O}(\lambda^{S_n+1}),
\nonumber 
\end{align}
The factor $P_n(w_1,\ldots,w_n)$ is the polynomial of the $n$-electron bound state,
which is also proportional to the FQH polynomial of $n$ electrons,
\begin{align}
 P_n(z_1,\ldots, z_n) \propto P(z_1,\ldots, z_n).
\end{align}
This factorization arises because, as a function of the $n$-cluster coordinate
$z^{(n)}$, all zeros are located at the electron coordinates $z_i$; the absence
of off-particle zeros is the defining property of the $n$-cluster
structure~\cite{WW0808,WW0809}.  

We may likewise bring $2n$ electrons together and obtain the limiting
factorization
\begin{align}
&
P(z_{1},\ldots, z_{N}) = \lambda^{S_{2n}}\, P_{2n}(w_1,\ldots, w_{2n})\, \times
\\
&\qquad\qquad
P(z_{2n+1},\ldots, z_{N})
\prod_{i=2n+1}^N (z_i-z^{(2n)})^m
+
\mathcal{O}(\lambda^{S_{2n}+1}).
\nonumber
\end{align}
In this way, we obtain a sequence of finite polynomials $P_{n}(z_1,\ldots,
z_{n})$, $P_{2n}(z_1,\ldots, z_{2n})$, $P_{3n}(z_1,\ldots, z_{3n}), \ldots$, or equivalently
\begin{align}
\label{PnP2n}
P(z_1,\ldots, z_{n}),\ P(z_1,\ldots, z_{2n}),\ \ldots,
\end{align}
which describe the FQH polynomial $P(\{z_i\})$.  

We now factor out the Laughlin factor arising from fusion of $n$ particles in an $n$-cluster state.
Since this factor in terms of the fusion coordinate is $\prod_i (z^{(n)}-z_i)^m$, the exponent before
fusion will be $m/n$.  We define
\begin{align}
	P_{kn}(z_1,\ldots,z_{kn}) = G_{kn}(z_1,\ldots, z_{kn})\prod_{i<j} (z_i-z_j)^{m/n}
\end{align}
The sequence of polynomials $G_{kn}$ for $k=1,2,\ldots$ will define the \emph{generalized pattern-of-zeros}
(gPOZ).  Clearly, the gPOZ includes the POZ data $S_a$.

The polynomial $G_{2n}$ contains information from two of the leading
$n$-cluster terms.  However, it also may contain information from subleading
terms of a single $n$-cluster, which nevertheless make up $S_{2n}$ zeros across a
$2n$-electron bound state.  Likewise, $G_{3n}$ may contain terms that were
subleading within a $2n$-electron bound state, but which make up $S_{3n}$
zeros across a $3n$-bound state.  Note that these terms are allowed from the
$n$-cluster definition, since the Jastrow-like property only applies to the
leading order of zero at the $n$-cluster level.  

To describe the minimal angular momentum state that is characterized by this data,
we now add a stability condition, that \frmbox{Removal of any zero
in the ground state wavefunction (\ie reducing total angular momentum of the
ground state wavefunction) causes a change in the data that describes gPOZ.  }
The utility of this is as follows.  First, it ensures that the gPOZ provides a
complete characterization of the state.  There are no higher-order zeros
which are ``invisible'' to the gPOZ.  Second, as a corollary, it allows such a
state to be constrained by a local Hamiltonian which is only sensitive to the
lowest order of zero between a collection of particles.  We expect that adding
additional zeros will be of no cost to such a Hamiltonian, so we will need
a further condition to enforce incompressibility,
\frmbox{Bulk excitation operators which create
$O(1)$ zeros annihilate the state.}  This will be in contrast to operators
which add a macroscopic number of zeros, expanding a droplet on the order of
one orbital into a confining potential and thus costing finite energy. 
$O(1)$ addition of zeros contribute $(1/N)$th such a disturbance, which is
consistent with an energy cost going to zero in the thermodynaic limit.  
We will explore this reasoning further, as well as
the distinction between bulk and edge operators, in the next section.

As mentioned above, in order for the gPOZ to be useful from a practical
standpoint, we must be able to characterize it using finite data.  Just as the
$n$-cluster condition allowed us to construct the infinite POZ
using just the first $n$ terms $S_1,\ldots,S_n$, our hope is that the
gPOZ may also be constructed by just knowing the first few $P_{2n}$.  

With the Laughlin factor removed from $P_{kn}$, the $n$-cluster condition means that
\begin{align}
	&G_N(z_1,\ldots, z_N) = \\
	&\;\;\lambda^{S_{kn}} G_{kn}(z_1,\ldots, z_{kn}) G_\text{red} (z_{kn+1}\ldots z_{N}) + O(\lambda^{S_{kn}+1})
\nonumber 
\end{align} 
The function $G_N$ separates into two independent functions $G$ at leading order in $\lambda$, one for an integer number
of $n$-clusters and another for the rest of the particles.  Assuming $N$ is divisible by
$n$, $G_\text{red}$ must also form the leading polynomial for an integer number of $n$-clusters, $G_{N-kn}$.
This is required by the stability condition; if $G_\text{red}$ contained more terms just than this leading
polynomial, being independent of coordinates other than $z_{kn+1},\ldots, z_N$, we would be able to remove these
zeros without causing a change in the gPOZ.  We note also that this factorization can also be understood as
a unique fusion condition; the fusion of $n$ particles results in a product form of the function $G$.  More
generally, the decomposition into a function of $n$ particles and $N-n$ particles could contain a sum, so this
constraint is quite strong.  

This implies that from a larger term $G_N$ in the gPOZ, we may obtain a smaller one $G_{N'}$ by fusing an integer
number of $n$-clusters, $N-N'$, and taking the reduced polynomial $G_\text{red} =
G_{N'}$.  For example, from
\begin{align}
G(z_1,\ldots, z_{5n}) \sim  G(z_1,\ldots, z_{2n}) G(z_1,\ldots, z_{3n})
\end{align}
when we bring $z_1,\ldots, z_{2n}$ together, we obtain a reduction
\begin{align}
 G(z_1,\ldots, z_{5n}) \to G(z_1,\ldots, z_{2n}).
\end{align}
We can also obtain this via an intermediate step,
\begin{align}
\label{associativity}
 G(z_1,\ldots, z_{5n}) \to G(z_1,\ldots, z_{3n})
\to
 G(z_1,\ldots, z_{2n}).
\end{align}
All such routes must yield the same result.  This will have implications for
building up larger $G_N$ from smaller ones: all must produce this associative
behavior.  


It will turn out that associativity and the scaling condition, combined with
the incompressibility requirement which we will state more explicitly in the
next section, will provide this finite data for building the gPOZ.  This
finite data will be encapsulated by the central charge $c$, leading to a
$(\nu,h,c)$ classification for FQH states.

\section{From \lowercase{g}POZ to OPE}

We will now lay out the procedure for building larger $G_{kn}$ from smaller ones.  For simplicity, we
focus on 2-cluster states for most of the remainder of the paper.  All can be generalized to $n$-cluster
states, with a three-cluster example given in Appendix~\ref{Z3}.

Briefly, we go back to the polynomials $P_{2\ell }$ containing the Laughlin factor.  Expanding in two coordiantes $z_1$
about $z_2$, we have
\begin{align}
	P_{2 \ell }(z_1,\ldots,z_{2\ell }) = \sum_{k=0}^\infty (z_1-z_2)^{S_2+k} \tilde f_k(z_2,\ldots,z_{2\ell })
\end{align}
where $S_2$ is the leading order of zero from the POZ.  From the 2-cluster condition, we already know $\tilde f_0$.
Recall that in replacing $z_i$ with $z^{(2)} + \lambda w_i$, the $z^{(2)}$ dependence
of the wavefunction is encapsulated by a factor of $\prod_i (z^{(2)} - z_i)^m$.  In the Taylor expansion, we have $z_2$ in the role
of $z^{(2)}$, so the $z_2$ dependence of $f_0$ is $\tilde f_0 \propto \prod_i (z_2 - z_i)^m$.  The rest of $\tilde f_0$
must reduce to $P_{2\ell-2}$ by our stability condition of the previous section.

When we remove the Laughlin factor, we obtain the expression for $G_{2\ell}$,
\begin{align}
	G_{2 \ell }(z_1,\ldots,z_{2\ell }) = \sum_{k=0}^\infty (z_1-z_2)^{-2h+k} f_k(z_2,\ldots,z_{2\ell })
\end{align}
Here we have labeled $-2h = S_2 - m/n$.  
Note that $2h$ is not necessarily integer.

Because $S_2$ is the leading order of zero in the wavefunction and
therefore less than the generic Laughlin factor $m/n$,
the leading term in the expansion of $G$ is a pole.
The accompanying function $f_0$ is independent of $z_2$ since we have
removed the Laughlin factor, so by our previous arguments it must be equal to
$G_{2\ell-2}$.  Therefore we have
\begin{align}
	G_{2\ell}(z_1,\ldots,z_{2\ell}) = \frac{1}{(z_1-z_2)^{2h}}G_{2(\ell-1)}(z_3,\ldots,z_{2\ell}) \nonumber \\
	+ \sum_{k=1}^\infty (z_1-z_2)^{-2h+k} f_k(z2,\ldots,z_{2\ell})
\end{align}
The $k=0$ term is very informative.  First, it gives us one term of $G_{2\ell}$ explicitly in terms
of $G_{2(\ell-1)}$, which will be our goal for all terms.  Second, it contains a pole.  By associativity,
this pole must occur between all pairs of particles in some part of the wavefunction, not only within $G_{2(\ell-1)}$, 
but also between pairs across $\{z_1,z_2\}$ and $\{z_3,\ldots,z_{2\ell}\}$.  This means
that the $k>2h$ terms in the expansion must at minimum build Taylor expansions of $(z_2-z_i)^{-2h}$ for all $i$.
There may be other poles at $k<2h$ as well, which should also be replicated within $k>2h$.

In order to build larger $G_N$ from smaller ones, we would like to express the functions $f_k$ for 
as operators $\phi_k$ acting on the previous term in the gPOZ, 
\begin{align}
	f_k(z_2,\ldots,z_{2\ell}) = \phi_k(z_2) G_{2(\ell-1)}(z_3,\ldots, z_{2\ell})
\end{align}  
This will allow us to iteratively build the gPOZ by fixing one expansion.  We now set out to define these operators.

\subsection{Operators}

Finding $\phi_k$ that produce the associative behavior we desire is a difficult problem.  In many cases
we know what Taylor expansion they should produce, though inserting these terms into the appropriate part of
$G_{2(\ell-1)}$ using general operators is nontrivial.
We will take an alternative approach, which is to first define a collection of operators generating allowed
changes of the function.  Since we are working with holomorphic (or more accurately, meromorphic) functions, we will start with the Witt
algebra, which forms the generators of holomorphic diffeomorphisms.  These obey the commutation relations
\begin{align}
	[L_n,L_m] = (n-m)L_{m+n}
\end{align}
A typical differential representation is $L_n = z^{n+1}\partial_z$.  Such operators introduce a fixed origin at 0, though,
and we would like to preserve translation invariance of the wavefunction.  We rectify this by performing all transformations of
$z_j$ relative to another particle coordinate $z_i$.  The corresponding Witt algebra generators are
$(z_i-z_j)^{n+1}\partial_{z_j}$.  Furthermore, our operators should enact the same behavior relative to all other particles, \ie 
they should preserve the symmetry already present in the wavefunction.  We therefore sum over all the coordinates $z_j$,
$\sum_{j;\,j\neq i} (z_i-z_j)^{n+1}\partial_{z_j}$.  
These operators do note immediately enforce symmetry of $z_i$ with the rest of the coordinates $z_j$, which is something we will have to require
from a collection of these operators together. Since $z_i$ is not being acted upon, it may be introduced to the wavefunction
via these operators.  We will take this to be the definition of an operator at coordinate $z_i$, notated $\mathcal{L}_n(z_i)$.  

In this differential representation of the Witt algebra, we are also allowed to have a term which multiplies $\sum_{j;\,j\neq i}(z_i-z_j)^n$.
To obey the Witt commutation relations, this must come with a linear-in-$n$ coefficient.  The differential
operators are thus
\begin{align}
	\mathcal{L}_n(z_i) = \sum_{j;\,j\neq i} -(z_j-z_i)^{n+1}\partial_{z_j}+ \lambda_n (z_j-z_i)^n
	\label{generalL}
\end{align}
for some $\lambda_n$ which we will soon fix.  Sometimes it will be important to specify the set of $z_j$
being summed over, in which cases we use the notation $\mathcal{L}_n(z_1;\{ z_j\})$.  In general, this
set of $z_j$ will be all coordinates appearing in the function to the right of the operator, except for
$z_i$ if it appears.

Our goal will to build the operator expansion $\phi_k$ using combinations of these basic allowed operators,
which will allowed us to iteratively construct $G_N$. Before we proceed, however, we must also lay out more
details of the incompressibility condition from the previous section.  
If we have a collection of operators of the form of Eqn~\ref{generalL}, these can be used to generate
excitations of the wavefunction.  An application of $\mathcal{L}_n(\xi;\{z_i\})$ modifies wavefunction behavior
at the distinguished coordinate $\xi$, preserving the symmetry of the remaining electron coordinates
$\{ z_i\}$.  This can be viewed as an excitation at the coordinate $\xi$.  

As stated in the introduction, when we add more zeros to the wave function, we expect
that the energy of a local Hamiltonian is not increased.  Adding a macroscopic number of zeros
will cost energy by causing the droplet to expand into a confining potential by the order of an electron orbital.  
A submacroscopic addition of zeros will cause $(1/N)$th this disturbance, so we expect the
energy increase from the confining potential to be negligible in the thermodynamic limit.
The operators $\mathcal{L}_n(z)$ for $n>0$
add $n$ zeros to the wavefunction, so we expect that acting with these for $n<<N$ will cost no energy.  To compare,
we recall that the Laughlin part of the wavefunction supports quasihole excitations $\prod_{i=1}^N (\xi-z_i)$,
which add $N$ zeros where $N$ is the total number of particles.  These are finite energy excitations.

Let us denote the collection of operators $L_n$ with $n>0$ $L_+$, and the collection of $L_n$ with $n<0$ $L_-$.
When we act with an operator $\mathcal{L}_{-n}(\xi)$ in $L_-$, we create poles in the wavefunction that 
will not in general be canceled by the Laughlin factor.  
One thing we can do to cancel these singularities is to add a Laughlin quasihole factor $\prod_i (z_i-\xi)^n$.
Together, these operators add a macroscopic number of zeros, so this is a gapped excitation. 
Another option is to take $\xi \rightarrow \infty$, \ie take the operator coordinate outside the bulk, which will 
now describe an edge excitation.
The effect on the wavefunction can be expressed by expanding in $z_i \ll z_\infty$, for instance
\begin{align}
	\lim_{z_\infty \rightarrow \infty} z_\infty^{n} \sum_{i} \frac{-\partial_{z_i}}{(z_\infty - z_i)^{n-1}} \sim \sum_i z_i\partial_{z_i} + \ldots
\end{align}
Edge excitations appear in a differential representation of the Witt algebra that breaks translation invariance, $z^{n+1}\partial_z$,
which is allowed because the presence of an edge breaks translation invariance.  


The situation with $L_+$ is different.  Here, $\mathcal{L}_n(z_i)$ 
may generate $O(1)$ zeros in the bulk $(z_i-z_j)^n$ without involving a Laughlin quasihole
factor. Therefore we expect excitations generated by $\mathcal{L}_n$ to be gapless
in the thermodynamic limit.  
It is in general very difficult to rigorously diagnose whether a change in the FQH polynomial corresponds to a 
bulk excitation or an edge excitation, so we ultimately take this as an assumption.
The intuition is that additional angular momentum from $\mathcal{L}_n(z_i)$
enters as relative angular momentum between particles in the bulk, $(z_i-z_j)$. 
In contrast, angular momentum from $\lim_{\xi\rightarrow\infty}\mathcal{L}_{-n}(\xi)$ appeared with respect to the origin, $\sum_i z_i^n$.  
For now we take it as an assumption that a nontrivial action of $L_+$ corresponds to gapless bulk excitations.
Later, when we make the connection with conformal field theory, it will become more apparent that $\bar{\mathcal{L}}_{+n}(z_i)$ acting
at a field coordinate produces a new field that is mutually local with other fields, which we can then interpret
as a bulk excitation.

It will turn out that in order to satisfy this requirement that $L_+$ annihilates the state, we will need to 
consider not just the Witt algebra but its central extension, the Virasoro algebra.
\begin{align}
	[L_n, L_m] = (n-m)L_{n+m} + \tfrac{c}{12}(n^3-n)\delta_{m+n}
\end{align}
This corresponds to a projective represention where the composition of two holomorphic diffeomorphisms $e^{\epsilon L_n}$ and
$e^{\epsilon L_{-n}}$ includes an additional multiplicative factor 
$g(c,n)$ on the original Witt result $e^{\epsilon L_n + \epsilon L_{-n} +2n\epsilon^2 L_0 + \ldots}$.
For now $c$ may take any complex value, though we will find that $L_+$ will only annihilate the state at specific
real values of $c$.

The central extension modifies the behavior between $L_{+n}$ and $L_{-n}$, and we can no longer use a differential representation
as in Eqn~\ref{generalL} for both $L_+$ and $L_-$ simultaneously, though we may still use it for one of these.  The
other will become a collection of more abstract operators.  These still encode much of the behavior
of the original Witt algebra, and we can still think of them as acting at a coordinate $z_i$ and adding 
order $n$ zeros to the wavefunction.  We will denote these $\bar{\mathcal{L}}_n(z_i)$, defined via
\begin{align}
	[\bar{\mathcal{L}}_n(z_i), \mathcal{L}_m(z_i)] = (n-m)L_{n+m} + \tfrac{c}{12}(n^3-n)\delta_{m+n}
\end{align}
We will end up using the concrete realization $\mathcal{L}_n(z)$ for the $n<0$ half of the algebra and 
the more abstract $\bar{\mathcal{L}}_n(z)$ for the $n>0$ half of the algebra.

We are now in a position to fix the coefficients $\lambda_n$ in the $\mathcal{L}_n(z)$.
First, note that we expect the algebra of allowed operators to include translations.  Concretely, if we have a function
$G_{2\ell}(z_1,\ldots,z_{2\ell})$, we should be able to shift the coordinate $z_1$ to $z_1'$ and retain
the same function.  This looks like a Taylor expansion,
\begin{align}
	G_{2\ell}(z_1',z_2,\ldots,z_{2\ell}) = \sum_{k=0}^{\infty} (z_1'-z_1)^k \frac{1}{k!} \partial_{z_1}^k G_{2\ell}(z_1,\ldots,z_{2\ell})
	\label{shift}
\end{align}
From translation invariance, $\sum_{j\neq i} -\partial_{z_j} = \partial_{z_i}$, so the operator
corresponding to $\partial_{z_1}$ is $\mathcal{L}_{-1}(z_1)$ with $\lambda_{-1} = 0$.

We now require the Taylor expansion of Eq.~\ref{shift} to be annihilated by $L_+$.  Here we must be careful about the
coordinates being acted upon. When we apply $\bar{\mathcal{L}}_{n}(z_1)$ to the whole expression,
it acts not only on $\{z_2,\ldots,z_{2\ell}\}$ to the right of the $\mathcal{L}_{-1}(z_1)$, but also on 
$(z_1'-z_1)^k$ to the left of $\mathcal{L}_{-1}(z_1)$.  
\begin{align}
	&\bar{\mathcal{L}}_{+n}(z_1;z_1',z_2,\ldots,z_{2\ell}) G_{2\ell}(z_1',z_2,\ldots,z_{2\ell}) \nonumber \\
	&= \bar{\mathcal{L}}_{+n}(z_1;z_1',z_2,\ldots,z_{2\ell})\sum_{k=0}^{\infty} (z_1'-z_1)^k \nonumber \\
	&\;\;\;\; \times \frac{1}{k!} \mathcal{L}_{-1}^k(z_1;z_2,\ldots z_{2\ell}) G_{2\ell}(z_1,\ldots,z_{2\ell})\nonumber \\
	&= 0
\end{align}
We now separate $\bar{\mathcal L}_{+n}$ into the piece acting on $z_1'$ and the piece acting on $\{z_2,\ldots, z_{2\ell}\}$.
The piece acting on $\{z_2,\ldots, z_{2\ell}\}$ is straightforward to deal with; it commutes 
through the $\mathcal{L}_{-1}$ according to the Virasoro algebra and annihilates $G$ on the right. 
(Note that these commutators do not involve $c$ yet, as $n^3-n=0$ for $n=1$).  Any further $\bar{\mathcal{L}}_{+m}$ 
generated by this procedure also $G$ on the right, so 
by this process we may reduce this to an expression only involving $\mathcal{L}_{-1}$ and $\mathcal{L}_0$.

The action on $z_1'$ does require a more concrete realization of $\bar{\mathcal{L}}_{+n}$.  This part of the expansion does not involve
any $\mathcal{L}_{-n}$, however.  Since the state is annihilated at this stage, it will also not at any later point involve $\mathcal{L}_{-n}$.
Therefore we are allowed to define the action of $\bar{\mathcal{L}}_n$ here using the differential operator $\mathcal{L}_{n}(z_1;z_1')$.

We have defined $\bar{\mathcal{L}}_{+n}(z_1;z_1',z_2,\ldots)$ to act on this particular expansion 
via $\mathcal{L}_{+n}(z_1;z_1') + \bar{\mathcal{L}}_{+n}(z_1;z_2,\ldots z_{2\ell})$,
Requiring this to annihilate the function, we obtain the condition
\begin{align}
	&\mathcal{L}_{+n}(z_1;z_1')\sum_{k=0}^{\infty} (z_1'-z_1)^k \frac{1}{k!} \nonumber \\
	& \quad \quad \times \mathcal{L}_{-1}^k(z_1;z_2,\ldots z_{2\ell})G_{2\ell}(z_1,\ldots,z_{2\ell})\nonumber \\
	&= -\bar{\mathcal{L}}_{+n}(z_1;z_2,\ldots, z_{2\ell})\sum_{k=0}^{\infty} (z_1'-z_1)^k \frac{1}{k!} \nonumber \\
	& \quad \quad \times \mathcal{L}_{-1}^k(z_1;z_2,\ldots z_{2\ell}) G_{2\ell}(z_1,\ldots,z_{2\ell})
	\label{lambdan}
\end{align}
As noted before, the right hand side of this expression can be reduced to an expression involving only $\mathcal{L}_{-1}(z_1)$
and $\mathcal{L}_0(z_1)$.  In general, functions that are annihilated by $L_+$ may be eigenfunctions of $L_0$ by the structure
of the algebra, $[L_\pm,L_0\oplus L_\pm] = L_\pm$.  If $G$ is such an eigenfunction, then Eq.~\ref{lambdan} reduces
to the condition that $\lambda_n = -(n+1)h$.  Let us now explore the implications of $G$ being an eigenfunction of $L_0$.

The explicit form of $\mathcal L_0$ is
\begin{align}
	\mathcal L_0(z_i) = \sum_{j;j\neq i} \left(-(z_j-z_i)\partial_{z_j} - h\right)
\end{align}
We assume by translation invariance that $G$ is composed of differences between particles, 
\begin{align}
	G = \sum_n c_n \prod_{i,j}(z_i-z_j)^{a^n_{ij}}
\end{align}
for some collection of exponents $a^n_{ij}$ and constants $c_n$.
There are two types of terms generated by $\mathcal L_0(z_i)$ acting on such a function.  One is from $(z_k - z_\ell)$ with $k,\ell \neq i$,
\begin{align}
	[\mathcal{L}_0(z_i), (z_k - z_\ell)^{a_{k\ell}}] = \left(\frac{(z_i-z_k)}{(z_k-z_\ell)} - \frac{(z_i-z_\ell)}{(z_k-z_\ell)}\right) \nonumber \\
	\times a_{k\ell} (z_k-z_\ell)^{a_{k\ell}^n} \nonumber\\
	= -a^n_{k\ell} (z_k-z_\ell)^{a^n_{k\ell}}
\end{align}
The other is from terms involving the distinguished coordinate $z_i$,
\begin{align}
	[\mathcal{L}_0(z_i;z_\ell), (z_i - z_\ell)^{a_{i\ell}}] = \frac{(z_i-z_\ell)}{(z_i-z_\ell)} a_{i\ell}(z_i-z_\ell)^{a^n_{ij}} \nonumber\\
	= a_{i\ell} (z_i-z_\ell)^{a^n_{ij}}
\end{align}
Now acting on the entire function, we find
\begin{align}
	&\mathcal L_0(z_i;\{z_j\}) G_{N}(z_i,\{z_j\}) = -h(N-1) +\nonumber \\
	& \quad \sum_n c_n \left(\sum_{k,\ell} -a^n_{k\ell} + \sum_\ell a^n_{i\ell}\right)\prod_{i,j}(z_i-z_j)^{a^n_{ij}}
\end{align}
For $G_N$ to be an eigenfunction with eigenvalue $h$, we must have that $h(N-1) - \sum_{k,\ell\neq i} a^n_{k\ell} + \sum_\ell a^n_{i\ell} = h$ for each component multiplying $c_n$.
One can check that this is the case for arbitrary $z_i$ if each particle $i$ has 
\begin{align}
	\sum_j a^n_{ij} = -2h
	\label{hzeros}
\end{align}
\ie, there are $2h$ poles involving each coordinate in the wavefunction.  If this is true, removing any zero will result in a collection of particles with more than
$2h$ poles per particle, and this will change the gPOZ, violating the stability condition.

We have now fixed the form of the operators $\mathcal{L}_n(z_i)$ acting on $G$ an eigenfunction of $L_0$ with eigenvalue $h$,
\begin{align}
	\mathcal{L}_n(z_i) = \sum_{j;\,j\neq i} -(z_j-z_i)^{n+1}\partial_{z_j} - h(n+1) (z_j-z_i)^n
	\label{Ln}
\end{align}

\subsection{Operator expansion}

We are now ready to fix the form of the operator expansion, which again is given by
\begin{align}
	&G_{2\ell}(z_1,\ldots,z_{2\ell}) \nonumber\\ 
	&= \sum_{k=0}^\infty (z_1-z_2)^{-2h+k} \phi_k(z_2) G_{2(\ell-1)}(z_3,\ldots,z_{2\ell})
	\label{expansion}
\end{align}
Note that the leading exponent $-2h$ is consistent with $G$ having an eigenvalue $h$ under
$L_0$, since particles $1$ and $2$ already satisfy Eq~\ref{hzeros} in the $k=0$ term.

The exponent $k$ on $(z_1-z_2)^{-2h+k}$ lowers the eigenvalue of $\mathcal{L}_0$ by
$k$, so $\phi_k$ must raise it again.  Each $\mathcal{L}_{-n}$ raises the
eigenvalue by $n$, so $\prod_i \mathcal{L}_{-n_i}$ with $\sum_i n_i = k$ form a
basis for $\phi_k$.  We may therefore write
\begin{align}
	\phi_k(z_2) = \sum_{n_1 + \ldots + n_i = k} \beta^{n_1,\ldots, n_i}\mathcal{L}_{-n_1}(z_2)\ldots \mathcal{L}_{-n_i}(z_2)
\end{align}
We can fix the $\beta$ by again applying $L_{+n}$ and requiring it to annihilate the expression in Eqn~\ref{expansion}.  Again, $L_+$ acts via 
$\mathcal{L}_n(z_2;z_1) + \bar{\mathcal{L}}_n(z_2;z_3,\ldots z_{2\ell})$.  The first term acts differentially on $(z_1-z_2)^{-2h+k}$, producing
a factor $2h-k+\lambda_n = -h(n-1)-k$ and raising the exponent by $n$.
The second term, $\bar{\mathcal{L}}$, commutes through the $\mathcal{L}_{-m_i}$ until annihilating $G_{2(\ell-1)}$ on the right, this time producing factors of $c$.  
Matching powers of $(z_1-z_2)$ on both sides, the coefficients in our operator series must therefore satisfy
\begin{align}
(h(n - 1) + k)\sum_{\substack{m_1+m_2+\ldots\\=k}}\beta^{\{m_i\}}\prod_i \mathcal{L}_{-m_i}(z_2) G_{2(\ell-1)} \nonumber \\
 = \sum_{\substack{n_1+n_2+\ldots\\=k-n}}\beta^{\{n_i\}}\bar{\mathcal{L}}_n(z_2) \prod_i \mathcal{L}_{-n_i}(z_2) G_{2(\ell-1)}
	\label{coeffs}
\end{align}
The $\beta$ at level $n+k$ can be found from those at level $k$ by commuting $L_{-m_i}$ so that the bases match on both sides.  
For example, applying $\bar{\mathcal{L}}_2(z_2)$ and examining the terms at order $k=0$, we find
\begin{align}
	h\beta^{\{0\}} G_{2(\ell-1)} = \bar{\mathcal{L}}_{2}(z_2;z_3,\ldots) \beta^{\{2\}}\mathcal{L}_{-2}(z_2;z_3,\ldots)  \nonumber\\ 
	\qquad \times G_{2(\ell-1)}(z_3,\ldots)\\
	h \beta^{\{0\}} G_{2(\ell-1)} = \beta^{\{2\}} (4\mathcal{L}_0(z_2;z_3,\ldots) + \tfrac{c}{12}(2^3-2)) \nonumber\\
	\qquad \times G_{2(\ell-1)}(z_3,\ldots)
\end{align}
From our earlier assertion that $\phi_0=1$, we have $\beta^{\{0\}} = 1$.
The $\mathcal{L}_0$ term on the right does not actually produce anything because $z_2$ does not appear in the function $G_{2(\ell-1)}$.
(This can be seen from our earlier discussion of how $\mathcal{L}_0$ acts on such terms; these contributions are all canceled by the
constant $hN$.)  We are then left with
\begin{align}
 	\beta^{\{2\}} =\tfrac{2h}{c}
\end{align}

We have now built an operator series that builds eigenfunctions of $\mathcal{L}_0$ that are annihilated by
$\mathcal{L}_+$.  There is no guarantee yet that this series is associative, \ie that it will converge to 
a function that manifests symmetry under fusing different subsets of particles in different orders.  This we will enforce in the next
section.  The only free parameter in this construction is the central charge, $c$, and we will find that 
associativity emerges only for specific values of $c$.

The central extension causes these generators of gapless excitations to take on an abstract form.
Since the wavefunction is built from the $L_-$ half of the algebra and the gapless
excitations originate from the $L_+$ half of the algebra, each of which separately close under
commutators, the behavior we require of $\bar{\mathcal{L}}_{+n}$ may seem somewhat arbitrary.
However, if we create an excitation at $\xi$ with $\mathcal{L}_{-n}(\xi)$, we expect the theory
to contain an operator which annihilates the excitation.  The operators $\bar{\mathcal{L}}_{+n}$
provide this operation, since
\begin{align}
	\bar{\mathcal{L}}_{+n}(\xi)\mathcal{L}_{-n}(\xi) G_N(\{ z_i\}) \\
	= \left(2n\mathcal{L}_0(\xi) + \frac{c}{12}(n^3-n)\right)G_N(\{ z_i\})
\end{align}
If $\xi$ is already a coordinate of $G_N$, then $\mathcal{L}_0(\xi)$ returns the eigenvalue
$h$.  If $\xi$ is not already a coordinate of $G_N$, which is a situation we consider,
then the right hand side of this expression vanishes unless there is a nonzero central
charge $c$.  Thus to be able to annihilate arbitrary excitations generated by $L_-$ without
annihilating the entire state, we must include $L_+$ in our theory with a central extension.

At this point, we also make the observation that this operator expansion exactly matches an operator product expansion
(OPE) in conformal field theory of two primary fields that fuse to the vacuum.  
Specifically, if $G_{n}$ is read as the $n$-point function $\< \phi(z_1)\ldots\phi(z_n)\>$ with $\phi \times \phi \sim \mathbb{I}$, 
Eq.~\ref{expansion} matches the OPE upon fusing $\phi(z_1)$ and $\phi(z_2)$.  This OPE is usually derived invoking
conformal invariance, but here we have arrived at the same machinery using assumptions which are natural to
quantum Hall states.  The key ingredient is the property that $L_+$ annihilates the function, which is in fact a defining
feature of primary fields of conformal field theory.  Here we have arrived at this feature via the incompressibility
of the quantum Hall state, providing an argument for why CFT correlators of primary fields should describe these states.

A possible alternate viewpoint which connects more concretely to traditional CFT is the observation that a correlator
of primary fields transforms under a global holomorphic diffeomorphism $z \rightarrow f(z)$ as $G(\{ z_i \}) \rightarrow
\prod_i (f'(z_i))^{h} G(\{ f(z_i)\})$.  In other words, the translation-invariant bulk function $G$ is preserved and the
only modification is by single-coordinate functions $f'(z)$ which may be viewed as edge states.  Correlators $G'$ of descendent fields,
on the other hand, are obtained by acting on $G$ with $\mathcal{L}_{-n}$, so when these functions transform they will involve
not only the original bulk $G'$ but also factors involving $G$ where the $\mathcal{L}_{-n}$ instead act on the single-coordinate functions.
This viewpoint may clearer in its distinction between bulk and edge, though it is more difficult to make a statement
incompressibility from transforming all of the particles at once.

\section{Determining the central charge}

We will enforce associativity by considering the most general form of a four-particle
function that satisfies the conditions of associativity and stability.  Such a function
will contain a parametrization
\begin{align}
	G(z_1,z_2,z_3,z_4) = \sum_n C_n \prod_{i<j} (z_i-z_j)^{a^n_{ij}}
\end{align}
with $C_n$ multiplying a linearly independent basis of allowed functions.  We have
written these functions in terms of $(z_i-z_j)$ to manifest translation invariance.
The exponenents $a^n_{ij}$ will satisfy certain requirements.  

First, because they are eigenfunctions of $L_0$, we must have
\begin{align}
	\sum_i a_{ij}^n = -2h
\end{align}
This ensures the correct number of poles involving each particle.  

Second, the function $G$ must have symmetry under particle exchange.  This requirement
will enforce associativity, since all pairs of particles are on equal footing.  The
specific symmetry will depend on $h$; for instance, if $2h$ is an even integer the
function will be symmetric under particle exchange, and if $2h$ is an odd integer the
function will be anti-symmetric under particle exchange.

Third, the highest-order singularity that can occur in $G$ is $(z_i-z_j)^{-2h}$.  To enforce
this, we will factor out a term containing all such singularities and consider a set of polynomials
$P_n$ (containing no poles) multiplying this.
\begin{align}
	&G(z_1,z_2,z_3,z_4) = \tilde G(z_1,z_2,z_3,z_4) \sum_n C_n  P_n(z_1,z_2,z_3,z_4) \\
	&\tilde G(\{ z_i\}) = \frac{1}{\prod_{i<j} (z_i-z_j)^{2h}}
\end{align}
To enforce our second condition, noticing that the correct symmetry under particle exchange
is contained in the factor $\tilde G$, we require that $P$ is a symmetric polynomial.
\begin{align}
	P_n(\{ z_i \}) = \mathcal{S}\left(\prod_{i<j} (z_i-z_j)^{b^n_{ij}} \right)
	\label{Pn}
\end{align}
For the four-particle function, since $\tilde G$ contains $\sum_{i\neq j} 2h = 6h$ poles for
particle $i$, $4h$ of them must be cancelled by $P$ in order to satisfy the first condition.
So we have for the exponents $b_{ij}^n$,
\begin{align}
	\sum_i b_{ij}^n = 4h
\end{align}
This formultion also allows for fractional $h$ more general than half-integer.
To see this, note that the factor $\tilde G$ will be exactly cancelled when we include the Laughlin factor 
$\prod_{i<j} (z_i-z_j)^m$ back in.  As long as we are careful about the ordering $i<j$ in the differences $(z_i-z_j)$,
this factor will be cancelled even for fractional exponents.  The condition on $h$ then is not that $2h$ be integer, 
but that the exponents $b_{ij}$ in $P_n$ be integer, since $P_n$ will enter the final wavefunction.  
Since $\sum_i b_{ij}^n = 4h$, this means that $h$ must be quantized to quarter-integer values.

We now proceed by generating all possible $P_n$ which satisfy these requirements and matching expansions 
of this parametrized four particle function $G$ with the operator expansion of the previous section. 
\begin{align}
&	G(z_1,\ldots,z_4) = \sum_{k=0}^\infty (z_1-z_2)^{-2h+k} \times
\nonumber\\
&\ \ \ \ \sum_{n_1+\ldots + n_i=k}\mathcal{L}_{-n_1}(z_2) \ldots \mathcal{L}_{-n_i}(z_2) (z_3-z_4)^{-2h}
\end{align}

To illustrate this procedure, we shall perform this proceedure on $h=3/2$, where the most general four-point function has
two terms:
\begin{align}
	G&(z_1,z_2,z_3,z_4) \nonumber \\
	&= \frac{1}{2}\mathcal{A}\left(\frac{1}{(z_1-z_2)^3(z_3-z_4)^3}\right) + \frac{C_1}{\prod_{i<j}(z_i-z_j)} 
\end{align}
Here $\mathcal{A}$ denotes antisymmetrization.  We have normalized so that the highest order pole appears
with coefficient 1, which automatically matches the operator expansion. The $k=1$ term in the expansion vanishes
by translation invariance ($\mathcal{L}_{-1} G_2 = 0$).  The next term in the expansion has
coefficient $\beta^{\{2\}} = 2h/c$.  Applying $\mathcal L_2$ to the two-point function, we generate a term
at order $(z_1-z_2)^{-1}$,
\begin{align}
	\beta^{\{2\}}&\mathcal{L}_2[z_2]\frac{1}{(z_3-z_4)^3}\nonumber\\
	&= \frac{9}{2 c (z_2 - z_3)^2 (z_2 - z_4)^2 (z_3 - z_4)}
\end{align}
This should match the $(z_1-z_2)^{-1}$ term in the expansion of $G$,
\begin{align}
	\frac{C_1}{(z_2 - z_3)^2 (z_2 - z_4)^2 (z_3 - z_4)}
\end{align}
We obtain $C_1 = 9/(2c)$.  

We now go to higher terms in the expansion to fix $c$.  The level 3 term
already matches without determining $c$, so we consider level 4.
The coefficients $\beta$ at this level are,
\begin{align}
	\beta^{\{4\}} =\frac{3 (4 + c - 2 h) h}{c (22 + 5 c)}, \quad \beta^{\{22\}} = \frac{2 h (1 + 5 h)}{c (22 + 5 c)}
\end{align}
Applying $\beta^{\{4\}}\mathcal L_4$, we get
\begin{align}
	\beta^{\{4\}}&\mathcal{L}_4[z_2]\frac{1}{(z_3-z_4)^3}=\left(\frac{9 + 9 c}{44 c + 10 c^2}\right)\times\nonumber\\
	&\frac{30 z_2^2 + 9 z_3^2 + 12 z_3 z_4 + 9 z_4^2 -  30 z_2 (z_3 + z_4)}{2 (z_2 - z_3)^4 (z_2 - z_4)^4 (z_3 - z_4)}
\end{align}
For $\beta^{\{22\}}\mathcal L_2 \mathcal L_2$, we have 
\begin{align}
	\beta^{\{22\}}&\mathcal{L}_{22}[z_2]\frac{1}{(z_3-z_4)^3}=\left( \frac{51}{44c+10c^2}\right)\times\nonumber\\
	&\frac{3 (12 z_2^2 + 7 z_3^2 - 2 z_3 z_4 + 7 z_4^2 - 12 z_2 (z_3 + z_4))}{4 (z_2 - z_3)^4 (z_2 - z_4)^4 (z_3 - z_4)}
\end{align}
Expanding $G$, we get at order $(z_1-z_2)$,
\begin{align}
	&\frac{3 (z_3 - z_4)}{(z_2 - z_3)^4 (z_2 - z_4)^4} + \nonumber\\
	&\;\;C_1\frac{(3 (3 z_2^2 + z_3^2 + z_3 z_4 + z_4^2 - 3 z_2 (z_3 + z_4))}{(z_2 - z_3)^4 (z_2 - z_4)^4 (z_3 - z_4)}
\end{align}
Equating these and substituting $C_1 = 9/(2c)$, we obtain the following equation for $c$
\begin{align}
 	40 c^2+ 182 c -147 =0
\end{align}
This quadratic equation has the two solutions $c=7/10$ and $c=-21/4$.  
$c=7/10$ corresponds to the $M_{5,4}$ minimal model.
We can verify that we have obtained the correct four-point function by means of 
a null vector equation, detailed in the Appendix.

Returning to general $h$, we present bases of linearly independent polynomials $P_n$ satisfying the requirements of translation invariance,
$\sum_i b_{ij} = 2h$ scaling, and symmetrization for quarter-integer $h$ up to $h=3$.  These are represented as tuples 
$(b_{12}, b_{34}, b_{13}, b_{24}, b_{14}, b_{23})$ of the exponents appearing in Eq.~\ref{Pn}.
\begin{align}
\begin{array}{|c|l|}
	\hline
	h & (b_{12}, b_{34}, b_{13}, b_{24}, b_{14}, b_{23}) \\
	\hline
	1/4 & \mathrm{none} \\
	1/2 & (0,0,1,1,1,1) \\
	3/4 & (0,0,1,1,2,2) \\
	1 & (0,0,2,2,2,2) \\
	5/4 & (0,0,2,2,3,3) \\
	3/2 & (0,0,3,3,3,3),\; (2,2,2,2,2,2) \\
	7/4 & (0,0,3,3,4,4) \\
	2 & (0,0,4,4,4,4), \; (2,2,3,3,3,3) \\
	9/4 & (0,0,4,4,5,5), \; (2,2,3,3,4,4) \\
	5/2 & (0,0,5,5,5,5), \; (2,2,4,4,4,4) \\
	11/4 & (0,0,5,5,6,6),\; (2,2,4,4,5,5) \\
	3 & (0,0,6,6,6,6),\; (2,2,5,5,5,5),\; (4,4,4,4,4,4) \\
	\hline
\end{array}
\end{align}
The first term in each row contains the leading pole.  This term is
normalized so that it matches the leading term in the operator expansion, with coefficient $C_0 = 1$.  
The subleading term(s), if present, introduce parametrization coefficients $C_n$ in 
the four-point function $G$.

Matching the operator expansion with expansions of these polynomials, we solve for any parametrization
coefficients and then find solutions for $c$.
In several cases, an initial equation for $c$ has multiple roots, but some of them fail
to match the operator expansion at higher order.  The solutions that give a convergent operator expansion are:
\begin{align}
\begin{array}{|c|c|c|}
	\hline
	h & c & \mathrm{minimal\ model}\\
	\hline
	\tfrac 14 & \mathrm{none} & \mathrm{none} \\
	\tfrac 12 & \tfrac 12 & M_{4,3}\\
	\tfrac 34 & -\tfrac 35 & M_{5,3} \\
	1 & 1 & \mathrm{Gaussian\ model} \\
	\tfrac 54 & -\tfrac {25}7 & M_{7,3} \\
	\tfrac 32 & \tfrac 7{10}, -\tfrac {21}4  & M_{5,4}, M_{8,3}\\
	\tfrac 74 & \mathrm{none} & \mathrm{none}\\
	2 & -\tfrac {44}5 & M_{10,3}\\
	\tfrac 94 & - \tfrac {117}{11} &  M_{11,3}\\
	\tfrac 52 & -\tfrac {13}{14} &  M_{7,4}\\
	\tfrac {11}4 & -\tfrac {187}{13} & M_{13,3}\\
	3 & \tfrac 45,-\tfrac {114}7 & M_{6,5}, M_{14,3}\\
	\hline
\end{array}
\end{align}
The solutions for $c$ all correspond to minimal models for CFT, or the chiral boson for $h=1$.  
Since the operator expansion we derived matches the OPE for two primary fields that fuse 
to the vacuum ($\Z_2$ simple currents),
having identified specific CFT models, we now know this expansion is associative
for any number of particles via associativity of the OPE.  

For the positive values of $c$, we have $h=1/2, c=1/2$ which is in the 
Ising CFT, $h=1,c=1$ which is in the free boson CFT, $h=3/2, c=7/10$ which is in the $M_{5,4}$ minimal model, 
and $h=3, c=4/5$ which is in the $M_{6,5}$ minimal model.  

The representation of $M_{5,4}$ in which the $\Z_2$ simple current is local is the $c=7/10$ superconformal
field theory.  This superconformal theory is equivalent to the Virasoro minimal model, with some of the
Virasoro primary fields packaged as superpartners via an additional $G_m$ component of the symmetry
algebra,
\begin{align}
	\{ G_m, G_n\} &\;= 2L_{m+n} + \tfrac{c}{3}(m^2-\tfrac{1}{4})\delta_{m+n}\nonumber \\
	[L_m,G_n] &\;= (\tfrac12 m-n)G_{m+n}
\end{align}
The $h=3/2$ field is the superpartner of the energy-momentum tensor.  We note that states built from
the ordinary energy-momentum tensor do not appear in our classification.  This is because $\mathcal{T}$
is not a primary field, and our approach relies on fields being annihilated by $L_+$.  The nonzero
contribution of $L_2\mathcal{T}$ would affect the coefficients in the operator expansion, so these would need to be
modified.
In contrast, the superpartner $G(z)$ is obtained by applying an element of $G_+$ to $\mathcal{T}(z)$.
$L_2$ now annihilates this combination, as well as $L_1$ because there is no first descendant of the 
vacuum. The state built from $G(z)$ is thus detected by our approach.

The $M_{6,5}$ model realization which manifests the $\Z_2$ simple current is the 3-state Potts model.
The $M_{6,5}$ model also has a $\Z_3$ parafermion realization, which has a $\mathbb{Z}_3$ simple current.
We will discuss the differences of these models in the next section.

We have also found negative solutions for $c$.  The simplest example is $h=3/4$
with $c=-3/5$, known as the Gaffnian state in the literature \cite{SB075317}.
In CFT, negative values of $c$ usually indicate the theory is non-unitary and
in particular has fields of negative norm.  In the context of a field theory,
negative norm indicates the theory is unphysical.  In the context of FQH
states, the bulk state has a norm taken as product of the holormorphic
correlator and its complex conjugate, so negative norm states do not appear
here.  However, the edge, viewed as a $1+1$d field theory with a spectrum
matching the bulk CFT, may manifest difficulty from the non-unitarity of the
CFT.  A possible resolution of this difficulty is that the bulk state may be
gapless.  Various numerical studies have tried to demonstrate the gaplessness
of the Gaffnian state \cite{JL075116,KM245117}.  Ref
\cite{FW045414} made an argument such a state must be gapless under
Hamiltonians obeying Lieb-Robinson bounds, which is the case for a screened
Coulomb interaction but may not be for a long-range Coulomb interaction.
Entanglement studies \cite{EB186801} indicate that the
correlation length of the Gaffnian state diverges in the thermodynamic limit,
so there is evidence of gaplessness.

In most of our examples, the many-body wavefunction becomes increasingly
complicated as the number of particles increases, so it is not practical to
explicitly write it down.  The threshhold where it does become possible to
compactly express an $N$-body wavefunction is $h \leq 1$ where the first
subleading term  in the operator expansion is regular.  These examples are well
known in the literature.  The examples for $h>1$ are more novel.

\section{Minimal models and quasiparticles}

Now we provide more details about the minimal models associated with the
$h=3/2$ and $h=3$ examples of the previous section.
This allows us to connect the ground states derived from our 2-cluster ansatz
with full theories containing other fields describing quasiparticles.  

We have identified our states with minimal models $M_{p,q}$ based on their central charge,
\begin{align}
	c = 1-6\frac{(p-q)^2}{pq}
\end{align}
The primary fields of $M_{p,q}$ are labeled by Kac indices $r,s \in \mathbb{Z}$ with
periodic identifications $\phi_{r,s} \sim \phi_{q-r,p-s} \sim \phi_{r+q,s+p}$.  Their 
conformal weights are
\begin{align}
	h_{r,s} = \frac{(pr - qs)^2 - (p-q)^2}{4pq}
\end{align}
and they have fusion rules
\begin{align}
	&\phi_{r_1,s_1} \times \phi_{r_2,s_2} = \nonumber \\
	&\sum_{r_3 \stackrel{\small 2}{=} |r_1-r_2|-1}^{\text{min}(r_1+r_2,2q-r_1-r_2)-1}\;
	\sum_{s_3 \stackrel{\small 2}{=} |s_1-s_2|-1}^{\text{min}(s_1+s_2,2p-s_1-s_2)-1} \phi_{r_3,s_3}
\end{align}
where the sums are in increments of $2$.  If $p$ or $q$ is even (we assume $q$ without loss
of generality) and at least 6, there is a subalgebra that is closed under fusion,
\begin{align}
	\mathcal{S}_D = \oplus_{r\stackrel{\small 2}{=} 1}^{q-1}\oplus_{s=1}^{p-1} \phi_{r,s}
\end{align}
This is called the D-series minimal model, while the one with all the Kac indices is called
the A-series minimal model.

The quantum dimension of a field $a$ is calculated
by constructing a matrix $N_{ab}^c$ with entries $0$ or $1$ depending on whether $a$ and $b$
fuse to $c$ and finding the largest eigenvalue of this matrix.

\subsection{$M_{5,3}$ realization of a $\Z_2$ simple current algebra}

The simplest 2-cluster state in our classification which hosts a Fibonacci anyon is the $M_{5,3}$ model, with $c=-3/5$ and $\Z_2$ simple current
$h=3/4$.  This theory has four primary fields,
\\[2mm]
\centerline{

\begin{tabular}{|c|c|c|c|c|c|c|}
\hline
primary fields& $\onebb$ 
& $J$ 
& $b$ 
& $c$ 
\\ 
\hline
$h$: & $0$
& $\frac{3}{4}$
& $\frac{1}{5}$
& $\frac{-1}{20}$
\\ 
\hline
$d$: & $1$
& $1$
& $\frac{1+\sqrt{5}}{2}$
& $\frac{1+\sqrt{5}}{2}$
\\ 
\hline
\end{tabular}
}
\vskip 2mm

The fusion is given by

\centerline{
\begin{tabular}{ |c||c|c|c|c|c|c|}
 \hline 
$\otimes$  & $\onebb$  & $J$ & $b$  & $c$ \\ 
\hline 
 \hline 
$\onebb$  & $ \onebb$  & $J$ & $b$ & $c$ \\
 \hline 
$J$  & $ J$ & $\onebb$ & $c$ & $b$ \\
 \hline 
$b$  & $ b$  & $c$ & $\onebb\oplus b$ & $a\oplus c$ \\
 \hline 
$c$  & $ c$  & $b$ & $a \oplus c$ & $\onebb\oplus b$ \\
 \hline 
\end{tabular}
}
\vskip 2mm

\subsection{$M_{5,4}$ realization of a $\Z_2$ simple current algebra ($SM_{3,5}$ superconformal field theory)}

The $M_{5,4}$ model contains
a $\Z_2$ simple current algebra $J^2 = \onebb$ with
\[
h(J)=\tfrac{3}{2},\qquad c=\tfrac{7}{10}.
\]
as found in the previous section.
This theory altogether has six primary fields:
\\[2mm]
\centerline{

\begin{tabular}{|c|c|c|c|c|c|c|}
\hline
primary fields& $\onebb$ 
& $J$ 
& $a$ 
& $b$ 
& $c$ 
& $d$ 
\\ 
\hline
$h$: & $0$
& $\frac{3}{2}$
& $\frac{7}{16}$
& $\frac{3}{5}$
& $\frac{1}{10}$
& $\frac{3}{80}$
\\ 
\hline
$d$: & $1$
& $1$
& $\sqrt{2}$
& $\frac{1+\sqrt{5}}{2}$
& $\frac{1+\sqrt{5}}{2}$
& $\frac{5+\sqrt{5}}{\sqrt{10}}$ 
\\ 
\hline
\end{tabular}
}
\vskip 2mm
Here $h$ is the conformal weight and $d$ is the quantum dimension.

The $S$-matrix follows standard minimal model structure 
\[\begin{pmatrix}
1,
& 1,
& \sqrt{2},
& \frac{1+\sqrt{5}}{2},
& \frac{1+\sqrt{5}}{2},
& \frac{5+\sqrt{5}}{\sqrt{10}} \\ 
1,
& 1,
& -\sqrt{2},
& \frac{1+\sqrt{5}}{2},
& \frac{1+\sqrt{5}}{2},
& \frac{-5-\sqrt{5}}{\sqrt{10}} \\ 
\sqrt{2},
& -\sqrt{2},
& 0,
& \frac{5+\sqrt{5}}{\sqrt{10}},
& \frac{-5-\sqrt{5}}{\sqrt{10}},
& 0 \\ 
\frac{1+\sqrt{5}}{2},
& \frac{1+\sqrt{5}}{2},
& \frac{5+\sqrt{5}}{\sqrt{10}},
& -1,
& -1,
& -\sqrt{2} \\ 
\frac{1+\sqrt{5}}{2},
& \frac{1+\sqrt{5}}{2},
& \frac{-5-\sqrt{5}}{\sqrt{10}},
& -1,
& -1,
& \sqrt{2} \\ 
\frac{5+\sqrt{5}}{\sqrt{10}},
& \frac{-5-\sqrt{5}}{\sqrt{10}},
& 0,
& -\sqrt{2},
& \sqrt{2},
& 0 \\ 
\end{pmatrix}
\]
The fusion channels are given by\\[2mm]
\centerline{
\begin{tabular}{ |c||c|c|c|c|c|c|}
 \hline 
$\otimes$  & $\onebb$  & $J$  & $a$  & $b$  & $c$  & $d$ \\ 
\hline 
 \hline 
$\onebb$  & $ \onebb$  & $ J$  & $ a$  & $ b$  & $ c$  & $ d$  \\ 
 \hline 
$J$  & $ J$  & $ \onebb$  & $ a$  & $ c$  & $ b$  & $ d$  \\ 
 \hline 
$a$  & $ a$  & $ a$  & $ \onebb \oplus J$  & $ d$  & $ d$  & $ b \oplus c$  \\ 
 \hline 
$b$  & $ b$  & $ c$  & $ d$  & $ \onebb \oplus b$  & $ J \oplus c$  & $ a \oplus d$  \\ 
 \hline 
$c$  & $ c$  & $ b$  & $ d$  & $ J \oplus c$  & $ \onebb \oplus b$  & $ a \oplus d$  \\ 
 \hline 
$d$  & $ d$  & $ d$  & $ b \oplus c$  & $ a \oplus d$  & $ a \oplus d$  & $ \onebb \oplus J \oplus b \oplus c$  \\ 
 \hline 
\end{tabular}
}
\vskip 2mm

The $M_{5,4}$ Virasoro minimal model may also be viewed as the superconformal
minimal model $SM_{3,5}$, where $SM_{p,q}$ has central charge~\cite{CR415402}
\begin{align}
	c = \frac{3}{2} - \frac{3(p-q)^2}{pq}
	\label{SCFTc}
\end{align}
and conformal dimensions
\begin{align}
	h_{r,s} = \frac{(pr - qs)^2-4}{8pq} + \frac{1}{32}(1-(-1)^{r-s})
\end{align}
Fields $\phi_{r,s}$ for $1\leq r \leq q-1$, $1 \leq s \leq s-1$ are in the Ramond sector
for $r-s\equiv 1\;\text{mod}\;2$ and in the Neveu-Schwarz sector for $r\equiv s \;\text{mod}\;2$.
The Neveu-Schwarz sector has half-integral $G_n$ in its basis, giving rise
to superpartners that differ by conformal weight $1/2$.  The field
$b$ becomes the superpartner of $c$, and $J$ is the superpartner of the energy-momentum tensor.
The other fields $a$ and $d$ lie in the Ramond sector, which has $G_n$ of integral
index in its basis.

The electron operator
\begin{align}
 \psi_e = J \, \ee^{\ii \alpha \phi}, \qquad \alpha=2,
\end{align}
produces a FQH state at
\[
\nu=\tfrac{1}{4},\qquad c=\tfrac{17}{10}.
\]
We refer to this as the $M_{5,4}$ state, supporting both Fibonacci and Ising non-Abelian statistics.

Representative quasiparticles include:
\begin{align}
 \psi_{q_1} &= c \, \ee^{\ii \beta \phi}, & \beta=1/2, &\quad Q_{q_1}=\tfrac{1}{4} &\quad d=\tfrac{1+\sqrt{5}}{2}, \\
 \psi_{q_2} &= b \, \ee^{\ii \beta \phi}, & \beta=1, &\quad Q_{q_2}=\tfrac{1}{2} &\quad d=\tfrac{1+\sqrt{5}}{2},\\%
 \psi_{q_3} &= a \, \ee^{\ii \beta \phi}, & \beta=\tfrac{3}{4}, &\quad Q_{q_2}=\tfrac{3}{8} &\quad d=\sqrt{2}.%
\end{align}
Here $\beta$ is determined to cancel the maximum pole order $h_J+h_M-h_{JM}$
for an anyon field $M$ with the simple current $J$.  $Q_q$ is the anyon charge $\alpha\beta/\alpha^2$.

The two Fibonacci quasiparticles $d=\frac{1+\sqrt{5}}{2}$ are superpartners
in the $SM_{3,5}$ superconformal field theory interpretation of this model, and together with $J$
they lie in the Neveu-Schwarz sector.  Since this sector is closed under the fusion algebra,
Fibonacci statistics may be realized by a model containing only this sector.  The Ising anyon 
lies in the other sector, \ie the Ramond sector.

The wavefunction for the minimally charged Fibonacci anyon at coordinate $\xi$ in the $M_{5,4}$ 
ground state is
\begin{align}
	\Psi_{q_1}(\xi,\{z_i\}) = \lim_{z_\infty \rightarrow \infty} z_\infty^{h_N} 
		\left\< \psi^N_{q_1}(z_\infty) \psi_{q_1}(\xi) \prod_i \psi_e(z_i) \right\>
\end{align}
where $\psi_{q_1}^N$ is $\psi_{q_1}e^{-iN\alpha\phi}$.  The term $z_\infty^{h_N}$ is 
fixed to balance the highest power of $z_\infty$ in the correlator.  We proceed
by calculating the $M_{5,4}$ factor for a single 2-cluster, $\< c(z_\infty) c(\xi) J(z_1) J(z_2) \>$.
The only fusion channel of $c\times c$ that contributes is the identity channel; the $b$ channel
does not lead to an overall vacuum expectation value.  Therefore we can straightforwardly obtain
\begin{align}
	&\< c(z_\infty) c(\xi) J(z_1) J(z_2) \> = \nonumber \\
	& \quad\sum_{\{n_i\}}\frac{\beta^{\{n_i\}}\prod_{n_i} \mathcal{L}_{n_i}[\xi; z_1,z_2]}{(z_\infty-\xi)^{1/5+\sum_i n_i}} \< J(z_1) J(z_2)\>
\end{align}
The series converges to
\begin{align}
	\< c(z_\infty) c(\xi) J(z_1) J(z_2) \> = \frac{1}{(z_\infty - \xi)^{1/5}(z_1-z_2)^3} \nonumber \\
	+\frac{3}{7} \frac{(z_\infty-\xi)^{9/5}}{(z_1-z_2)\prod_{i=1}^2(z_\infty- z_i)(\xi-z_i)}
\end{align}
Combining with the vertex operator part and taking the limit $z_\infty \rightarrow \infty$, we obtain 
the anyon wavefunction,
\begin{align}
	\Psi_{q_1}(\xi,\{ z_1,z_2\}) =  (z_1-z_2)\prod_{i=1}^2(\xi-z_i) + \tfrac{3}{7} (z_1-z_2)^3
\end{align}
We see that there is a term where the anyon is connected to each other
particle by $\alpha\beta$ zeros.  Such a term is at leading order in the operator expansion 
and is present in the thermodynamic limit.  There is a further term where the
anyon appears neutral.  In the thermodynamic limit, terms near this level will
cancel $O(1)$ zeros containing $\xi$.  The cancellation may continue all the
way up to $O(N)$ in which terms the anyon would appear neutral, but these will
appear at level $N$ in the operator expansion and be suppressed by a factor $1/N!$  from the
coefficient.  Therefore, we consider the quasiparticle to have charge
$\alpha\beta/\alpha^2$ in the thermodynamic limit.

\subsection{$M_{8,3}$ realization of a $\Z_2$ simple current algebra ($SM_{2,8}$ superconformal model)}

The other example of a $\Z_2$ simple current with $h=3/2$ that we found had
central charge $c=-21/4$.  This corresponds to a non-unitary Virasoro minimal
model $M_{8,3}$, which has seven primary fields.  We list their conformal
weights and Frobenius-Perron dimensions (FPdim) below. 
Here FPdims are
defined as the largest real eigenvalues as the fusion matrices, which
are not given by the unit row of the $S$-matrix for non-unitary case:
\vskip 1mm
\centerline{
\begin{tabular}{|c|c|c|c|c|c|c|c|}
\hline
primary fields& $\onebb$ 
& $J$ 
& $a$ 
& $b$ 
& $c$ 
& $d$ 
& $e$ 
\\ 
\hline
$h$: & $0$
& $3/2$
& $-\frac{7}{32}$
& $-\frac{1}{4}$
& $-\frac{3}{32}$
& $\frac{1}{4}$
& $\frac{25}{32}$
\\ 
\hline
FPdim: & $0$
& $1$
& \scalebox{.6}{$\sqrt{2 + \sqrt{2}}$}
& \scalebox{.6}{$1+\sqrt{2}$}
& \scalebox{.6}{$\sqrt{4 + 2\sqrt{2}}$}
& \scalebox{.6}{$1+\sqrt{2}$}
& \scalebox{.6}{$\sqrt{2 + \sqrt{2}}$}
\\ 
\hline
\end{tabular}
}
\vskip 1mm

The fusion rules are
\\[1mm]
\hfil
\begin{tabular}{ |c||c|c|c|c|c|c|}
 \hline 
$\otimes$  & $J$ & $a$  & $b$  & $c$  & $d$  & $e$ \\ 
\hline 
 \hline 
$J$  & $J$ & $ e$  & $ d$  & $ c$  & $ b$ & $a$  \\ 
 \hline 
$a$  & $e$ & $ \onebb \oplus b$  & $ a\oplus c$  & $ b\oplus d$  & $ c\oplus e$ & $d\oplus J$ \\ 
 \hline 
$b$  &$d$ &$a\oplus c$  &$\onebb\oplus b\oplus d$  &$a\oplus c\oplus e$ &$b\oplus d\oplus J$ &$c\oplus e$ \\ 
 \hline 
$c$ & $c$ & $ b\oplus d$  &$a\oplus c\oplus e$  &$\onebb\oplus b\oplus d\oplus J$ &$a\oplus c\oplus e $ &$b\oplus d$ \\ 
 \hline 
$d$  & $b$ & $ c\oplus e$  & $ b\oplus d\oplus J$  & $ a\oplus c\oplus e$  & $ \onebb \oplus b \oplus d$ &$a \oplus c$  \\ 
 \hline
$e$  & $a$ & $ d\oplus J$  & $c\oplus e$ & $ b\oplus d$  & $ a\oplus c $ & $\onebb\oplus b$  \\ 
 \hline 
\end{tabular}
\vskip 2mm

This minimal model also corresponds to superconformal theory $SM_{p,q}$. Here we find $q/p = 4$ 
gives the correct central charge, and we choose $q=8,p=2$ satisfying additional conditions 
$p\equiv\; q\; \text{mod} 2$ and $\text{gcd}(p,(q-p)/2)=1$~\cite{CR415402}.  The 
fields in the Ramond sector are $b$
with its superpartner field $d$, and $J$ which is the superpartner of $\mathcal{T}$.  In the Ramond
sector we have fields $a$, $c$, and $e$.

The FQH state occus at $\nu=1/4$ and has electron operator
\begin{align}
	\psi_e = J \ee^{\ii\alpha \phi}, \quad \alpha = 2
\end{align}
This has the same filling fraction and pattern of zeros as the $M_{5,4}$
state, but its behavior is vastly different.  This theory does not have
Fibonacci anyons.
We also see negative conformal weights, which is not a problem for our bulk
wavefunction where zeros are preferred over poles anyway, but this does pose a
concern for more standard CFT which may be present in edge states.  Exploring
the properties of this state is an interesting direction of future work.

\subsection{$M_{6,5}$ realization of a $\Z_2$ simple current algebra}

The $M_{6,5}$ minimal model with $c=4/5$ contains the $\Z_2$ simple current of
conformal weight $h=3$.  It has the $10$ Virasoro primaries, which are
labelled by $(r,s)$, $1\le r\le 4$, $1\le s\le 5$, modulo
$(r,s)\sim(5-r,6-s)$, with $h_{r,s}=\dfrac{(6r-5s)^2-1}{120}$. We fix the
ordered basis
\begin{widetext}
\hfil \begin{tabular}{|l|c|c|c|c|c|c|c|c|c|c|}
\hline
\# & $1=\onebb$ & 2 & $3=Z$ & 4 & $5=J$ & $6=W$ & 7 & $8=Y$ & 9 & $10=X$\\
\hline
Primary fields & $(1,1)$&$(1,2)$&$(1,3)$&$(1,4)$&$(1,5)$&$(2,1)$&$(2,2)$&$(2,3)$&$(2,4)$&$(2,5)$\\
\hline
$h$: & $0$ & $\tfrac18$ & $\tfrac23$ & $\tfrac{13}{8}$ & $3$ & $\tfrac25$ & $\tfrac1{40}$ & $\tfrac1{15}$ & $\tfrac{21}{40}$ & $\tfrac75$\\
\hline
$d$: & $1$ & $\sqrt3$ & $2$ & $\sqrt3$ & $1$ & $\varphi$ & $\sqrt3\varphi$ & $2\varphi$ & $\sqrt3\varphi$ & $\varphi$ \\
\hline
\end{tabular}
\end{widetext}
where $ \varphi=\tfrac{1+\sqrt5}{2}$.

The $S$-matrix is given by
\[
S_{(r,s),(r',s')}
=\frac{2}{\sqrt{15}}\,(-1)^{1+sr'+rs'}\,
\sin\!\Big(\tfrac{6\pi}{5}\,rr'\Big)\,
\sin\!\Big(\tfrac{5\pi}{6}\,ss'\Big),
\]
a real, symmetric, unitary $10\times10$ matrix. Only two independent radial
factors appear,
\[
K_1=\frac{2}{\sqrt{15}}\sin\tfrac{\pi}{5}=0.303,\quad
K_2=\frac{2}{\sqrt{15}}\sin\tfrac{2\pi}{5}=0.491 ,
\]
and all entries equal $\pm K_i,\ \pm\tfrac{\sqrt3}{2}K_i,\ \pm\tfrac12 K_i$ or $0$.
In particular $S_{00}=\tfrac12 K_1=\sin(\pi/5)/\sqrt{15}=0.151766$, and the total
quantum dimension is $\mathcal D=1/S_{00}=\sqrt{30+6\sqrt5}=6.5897$.

With the abbreviations
\begin{align}
&a_1=\tfrac12K_1,\quad a_2=\tfrac{\sqrt3}{2}K_1,\quad a_3=K_1,
\nonumber\\
&c_1=\tfrac12K_2,\quad c_2=\tfrac{\sqrt3}{2}K_2,\quad c_3=K_2,
\end{align}
the $S$-matrix (rows/columns in the order above) is
\[
\left(
\begin{array}{ccccc|ccccc}
 a_1 & a_2 & a_3 & a_2 & a_1 &  c_1 & c_2 & c_3 & c_2 & c_1\\
 a_2 &-a_2 & 0   & a_2 &-a_2 & -c_2 & c_2 & 0   &-c_2 & c_2\\
 a_3 & 0   &-a_3 & 0   & a_3 &  c_3 & 0   &-c_3 & 0   & c_3\\
 a_2 & a_2 & 0   &-a_2 &-a_2 & -c_2 &-c_2 & 0   & c_2 & c_2\\
 a_1 &-a_2 & a_3 &-a_2 & a_1 &  c_1 &-c_2 & c_3 &-c_2 & c_1\\ \hline
 c_1 &-c_2 & c_3 &-c_2 & c_1 & -a_1 & a_2 &-a_3 & a_2 &-a_1\\
 c_2 & c_2 & 0   &-c_2 &-c_2 &  a_2 & a_2 & 0   &-a_2 &-a_2\\
 c_3 & 0   &-c_3 & 0   & c_3 & -a_3 & 0   & a_3 & 0   &-a_3\\
 c_2 &-c_2 & 0   & c_2 &-c_2 &  a_2 &-a_2 & 0   & a_2 &-a_2\\
 c_1 & c_2 & c_3 & c_2 & c_1 & -a_1 &-a_2 &-a_3 &-a_2 &-a_1
\end{array}
\right).
\]

With $c/24=\tfrac{1}{30}$, the $T$-matrix is given by
\begin{align}
&T=\ee^{-\pi i/15}\,\operatorname{diag}\big(\ee^{2\pi i h_a}\big)
\nonumber\\
&=\ee^{-\pi i/15}\,\operatorname{diag}\!\Big(
1,\,\ee^{\ii\pi/4},\,\ee^{4\pi i/3},\,\ee^{5\pi i/4},\,1,\,
\ee^{4\pi i/5},
\nonumber\\
&
\ \ \ \  
\ \ \ \  
\ee^{\ii\pi/20},\,\ee^{2\pi i/15},\,\ee^{21\pi i/20},\,\ee^{4\pi i/5}\Big).
\end{align}
(Note $h_{1,5}=3$ and $h_{2,5}=\tfrac75\equiv\tfrac25\ (\mathrm{mod}\,1)$, so
the $5$th and $10$th phases coincide with the $1$st and $6$th respectively;
the boson $\phi_{1,5}$ appears as $\ee^{2\pi i\cdot3}=1$.)

The radial ($r$) index carries $\widehat{su}(2)_3$ data (quantum dimensions
$1,\varphi,\varphi,1$, i.e.\ a Fibonacci factor) and the ($s$) index carries
$\widehat{su}(2)_4$ data ($1,\sqrt3,2,\sqrt3,1$); the $\mathbb Z_2$
identification $(r,s)\sim(5-r,6-s)$ glues them into the $10$--anyon modular
category $\mathcal C_{(6,5)}$ used above.

Although there is an $h=2/3$ field in this theory, it is not the $\Z_3$ simple
current of the $\Z_3$ parafermion theory.  However, the $M_{6,5}$ model is
closely related to the $\Z_3$ parafermion CFT, which we will describe in the
next section \ref{Z3pf}.  

In fact, $J$ in $M_{6,5}$ model generates an condensible algebra
$A=\onebb\oplus J$, whose condensation reduces the $M_{6,5}$ minimal model to
the $\Z_3$ parafermion CFT.  The $J$-fusion is given by
$J\times(r,s)=(r,6-s)$.  The $\mathbb Z_2$ monodromy charge
$Q_J(a)=h_J+h_{(r,s)}-h_{J\times (r,s)}\bmod 1$ controls deconfinement:
\[
\begin{array}{|l|c|c|c|c|}
\hline
\text{orbit of $J$ fusion} & Q_J & \text{status} & \text{parafermion}\\[2pt]
\hline
\{(1,1)_\onebb,(1,5)_J\} & 0 & \text{deconfined} & \onebb\\
\{(2,1)_W,(2,5)_X\} & 0 & \text{deconfined} & M_0\\
(1,3)_Z\ \text{(fixed)} & 0 & \text{splits} & J_1\oplus J_2\\
(2,3)_Y\ \text{(fixed)} & 0 & \text{splits} & M_1\oplus M_2\\
\hline
\{(1,2),(1,4)\} & \tfrac12 & \text{confined} & \text{(twist sector)}\\
\{(2,2),(2,4)\} & \tfrac12 & \text{confined} & \text{(twist sector)}\\
\hline
\end{array}
\]
where $\onebb,J_1,J_2,M_0,M_1,M_2$ are the six primary fields of
the $\Z_3$ parafermion CFT.


The fusion of $\onebb,J,X,Y,Z$ are closed with fusion rule:\\[2mm]
\centerline{
\begin{tabular}{ |c||c|c|c|c|c|c|}
 \hline 
$\otimes$  & $\onebb$  & $J$  & $W$  & $X$  & $Y$  & $Z$ \\ 
\hline 
 \hline 
$\onebb$  & $ \onebb$  & $J$  & $W$  & $X$  & $Y$  & $Z$  \\ 
 \hline 
$J$  & $J$  & $ \onebb$  & $ X$  & $ W$  & $ Y$  & $ Z$  \\ 
 \hline 
$W$  & $W$  & $ X$  & $ \onebb \oplus X$  & $ W\oplus J$  & $ Y\oplus Z$  & $ Y$  \\ 
 \hline 
$X$  & $X$  & $ W$  & $ W\oplus J$  & $ \onebb \oplus X$  & $ Y \oplus Z$  & $ Y$  \\ 
 \hline 
$Y$  & $Y$  & $ Y$  & $ Y\oplus Z$  & $ Y \oplus Z$  & all & $ W \oplus X \oplus Y$  \\ 
 \hline 
$Z$  & $Z$  & $ Z$  & $ Y$  & $ Y$  & $ W\oplus X\oplus Y$  & $ \onebb \oplus J \oplus Z $  \\ 
 \hline 
\end{tabular}
}
\vskip 2mm
where ``all'' indicates fusion into all six channels $1 \oplus J \oplus W
\oplus X \oplus Y \oplus Z$.  From the $S$-matrix, we see that $S_{J,i} >0$
for $i \in \{\onebb,J,X,Y,Z,W\}$ and $S_{J,i} <0$ otherwise.  Thus
$\{\onebb,J,X,Y,Z,W\}$ are local respect to $J$, while correlation of $J$ with
$\{\phi_{(1,2)}, \phi_{(1,4)}, \phi_{(2,2)}, \phi_{(2,4)}\}$
has a $-1$ branch cut.

The electron operator
\begin{align} 
	\psi_e = J \ee^{\ii\alpha\phi}, \quad \alpha = \sqrt{7} 
\end{align}
produces a FQH state at $\nu = 1/7$ with central charge $c=4/5+1$.  This state
supports $d=\vphi$ Fibonacci statistics with representative quasiparticles
\begin{align}
	\psi_{q_1} = W \ee^{\ii\beta\phi}, \quad \beta = \frac{2}{\sqrt{7}}, \quad Q_q = \frac{2}{7}\\
	\psi_{q_2} = X \ee^{\ii\beta\phi}, \quad \beta = \frac{4}{\sqrt{7}}, \quad Q_q = \frac{4}{7}
\end{align}
as well as $d=\sqrt 3$ non-Abelian statistics  with representative quasiparticles
\begin{align}
	\psi_{q_3} = \phi_{(1,2)} \ee^{\ii\beta\phi}, \quad \beta = \frac{3}{2\sqrt{7}}, \quad Q_q = \frac{3}{14}\\
	\psi_{q_4} = \phi_{(1,4)} \ee^{\ii\beta\phi}, \quad \beta = \frac{9}{2\sqrt{7}}, \quad Q_q = \frac{9}{14}
\end{align}

The wavefunction for the minimally charged Fibonacci quasiparticle with two electrons is obtained from
\begin{align}
	\Psi_{q_1}(\xi,\{z_1,z_2\}) = \lim_{z_\infty\rightarrow\infty} z_\infty^{h_N}\< \psi^N_{q_1}(z_\infty) \psi_{q_1}(\xi) \psi_e (z_1) \psi_e(z_2)\>
\end{align}
which we calculate to be
\begin{align}
	\Psi_{q_1}(\xi,\{z_1,z_2\}) = (z_1-z_2) \prod_{i=1}^2 (\xi-z_i)^2 \nonumber \\
	+ 3 (z_1-z_2)^3 \prod_{i=1}^2 (\xi-z_i) + \tfrac{21}{26} (z_1-z_2)^5
\end{align}
Subleading zeros of the anyon appear with coefficient $3$ and $\tfrac{21}{26}$.

The other Fibonacci anyon has 
\begin{align}
	&\Psi_{q_2}(\xi,\{z_1,z_2\}) = (z_1-z_2) \prod_{i=1}^2 (\xi-z_i)^4 \nonumber \\
	&+ \tfrac{21}{2} (z_1-z_2)^3 \prod_{i=1}^2 (\xi-z_i)^2 + \tfrac{483}{52} (z_1-z_2)^5\prod_{i=1}^2 (\xi-z_i)^2 \nonumber \\
	&+ \tfrac{147}{26} (z_1-z_2)^7 \prod_{i=1}^2 (\xi-z_i) + \tfrac{21}{26} (z_1-z_2)^{9}
\end{align}

In this case, the ratio of $h$ to $c$ is such that the first few terms are larger than the
leading term.  However, we see already see the terms after this begin to decay, and the 
neutral anyon term will still be exponentially small in the thermodynamic limit.

\subsection{$\Z_3$ parafermion realization of a $\Z_3$ simple current
algebra}
\label{Z3pf}

The $Z_p$ para\-fermion CFT is the coset
\[
\frac{su(2)_p}{u(1)} \, .
\]
For $p=3$ one obtains the $Z_3$ parafermion theory.
The central charge is
\[
c \;=\; \frac{2(p-1)}{p+2}\, , \qquad \Rightarrow \qquad c\big|_{p=3} \;=\; \frac{4}{5}\, .
\]

The chiral algebra is generated by parafermion (simple) currents $J_r(z)$ with $r=1,\dots,p-1$ carrying $\Z_k$ charge $r$.
Their conformal weights are
\[
h(J_r) \;=\; \frac{r(p-r)}{p} \, .
\]
For $p=3$ there are two nontrivial chiral parafermions,
\[
h(J_1)=h(J_2)=\frac{2}{3}\,,
\]
with the $\Z_3$-graded fusion (and operator expansion) pattern
\[
J_r \times J_s \;\sim\; J_{\,r+s \,\text{mod}\,3}\,,\qquad J_0\equiv \mathbf{1}\, .
\]
Here and below, $\sim$ means equality up to descendant contributions and structure constants.

Primary fields of the coset are labeled by pairs
\[
\Phi^{\ell}_{m}\,,\qquad \ell=0,1,\dots,p,\qquad m\in\Z_{2p}\,,
\]
subject to the selection rule
\[
\ell+m\ \text{even}\,,
\]
and the field identification
\[
\Phi^{\ell}_{m}\ \equiv\ \Phi^{\,k-\ell}_{\,m+p}\,.
\]
Their holomorphic conformal weights are
\[
h(\Phi^{\ell}_{m}) \;=\; \frac{\ell(\ell+2)}{4(p+2)} \;-\; \frac{m^2}{4p}\,,
\qquad (\text{with }\ell+m\text{ even})\, .
\]

Let $N^{\ell}_{\ell_1,\ell_2}$ denote the $su(2)_p$ fusion coefficients: 
\begin{align}
& \ell_1 \otimes \ell_2 \; =\; \bigoplus_{\ell}
N^{\ell}_{\ell_1,\ell_2}\, \ell\, ,
\\
\ell &= |\ell_1-\ell_2|,|\ell_1-\ell_2|+2,\dots,\min(\ell_1+\ell_2,\,2k-\ell_1-\ell_2) .
\nonumber 
\end{align}
Then the parafermion fusion is
\[
\Phi^{\ell_1}_{m_1}\ \times\ \Phi^{\ell_2}_{m_2}
\;=\;
\bigoplus_{\ell}\,
N^{\ell}_{\ell_1,\ell_2}\;\Phi^{\ell}_{\,m_1+m_2\ (\text{mod }2p)}
\]
with the understanding that (i) one keeps only summands obeying the selection rule $\ell+(m_1+m_2)$ even, and (ii) fields related by the identification
$\Phi^{\ell}_{m}\equiv \Phi^{p-\ell}_{m+p}$ are the same primary.


The $\Z_3$ parafermion CFT
has six primaries (mod selection rule $\ell+m$ even and identification $(\ell,m)\sim(3-\ell,m+3)$), with the following conformal weights and quantum dimensions:
\[
\begin{array}{|c|c|c|c|}
\hline
\text{primary field} & (\ell,m) & h & d \\
\hline
\mathbf{1} & (0,0) & 0 & 1 \\
J_1 & (0,2) & \frac23& 1 \\
J_2 & (0,4) & \frac23& 1 \\
M_0 & (2,0) & \frac 25& \frac{1+\sqrt 5}{2} \\  
M_1 & (1,1) & \frac1{15}& \frac{1+\sqrt 5}{2} \\
M_2 & (1,5)\equiv(1,-1) & \frac1{15}& \frac{1+\sqrt 5}{2} \\
\hline
\end{array}
\]
Here $J_1$ is an order-3 simple current.
The modular $S$-matrix is given by
\begin{align}
S=\begin{pmatrix}
1,
& 1,
& 1,
& \vphi,
& \vphi,
& \vphi \\ 
1,
& -\zeta_{6}^{1},
& \zeta_{3}^{1},
& \vphi,
& -\vphi\zeta_{6}^{1},
& \vphi\zeta_{3}^{1} \\ 
1,
& \zeta_{3}^{1},
& -\zeta_{6}^{1},
& \vphi,
& \vphi\zeta_{3}^{1},
& -\vphi\zeta_{6}^{1} \\ 
\vphi,
& \vphi,
& \vphi,
& -1,
& -1,
& -1 \\ 
\vphi,
& -\vphi\zeta_{6}^{1},
& \vphi\zeta_{3}^{1},
& -1,
& \zeta_{6}^{1},
& -\zeta_{3}^{1} \\ 
\vphi,
& \vphi\zeta_{3}^{1},
& -\vphi\zeta_{6}^{1},
& -1,
& -\zeta_{3}^{1},
& \zeta_{6}^{1} \\ 
\end{pmatrix}
\end{align}
where
\begin{align}
&
\zeta^m_n=\mathrm{e}^{2\pi \mathrm{i} m/n},\ \ \
\vphi=\frac{1+\sqrt 5}{2}.
\end{align}

The fusion rule is given by\\[2mm]
\centerline{
\begin{tabular}{ |c||c|c|c|c|c|c|}
 \hline 
$\otimes$  & $\onebb$  & $J_1$  & $J_2$  & $M_0$  & $M_1$  & $M_2$ \\ 
\hline 
 \hline 
$\onebb$  & $ \onebb$  & $ J_1$  & $ J_2$  & $ M_0$  & $ M_1$  & $ M_2$  \\ 
 \hline 
$J_1$  & $ J_1$  & $ J_2$  & $ \onebb$  & $ M_1$  & $ M_2$  & $ M_0$  \\ 
 \hline 
$J_2$  & $ J_2$  & $ \onebb$  & $ J_1$  & $ M_2$  & $ M_0$  & $ M_1$  \\ 
 \hline 
$M_0$  & $ M_0$  & $ M_1$  & $ M_2$  & $ \onebb \oplus M_0$  & $ J_1 \oplus M_1$  & $ J_2 \oplus M_2$  \\ 
 \hline 
$M_1$  & $ M_1$  & $ M_2$  & $ M_0$  & $ J_1 \oplus M_1$  & $ J_2 \oplus M_2$  & $ \onebb \oplus M_0$  \\ 
 \hline 
$M_2$  & $ M_2$  & $ M_0$  & $ M_1$  & $ J_2 \oplus M_2$  & $ \onebb \oplus M_0$  & $ J_1 \oplus M_1$  \\ 
 \hline 
\end{tabular}
}
\vskip 2mm

Similar to how the $\Z_3$ simple current was not apparent in the $M_{6,5}$
minimal model, the $\Z_2$ simple current is not apparent here.  The two CFTs
are relateded by a condensation of $J$ in $M_{6,5}$ minimal model, which
reduces $M_{6,5}$ CFT to $\Z_3$ parafermion CFT.

Using the $\Z_3$ parafermion current operator $J_1$, we can construct the
electron operator
\begin{align}
 \psi_e =J_1 \ee^{\ii \al \phi}, \ \ \ \al=\sqrt{\frac23+1}
\end{align}
which leads to a FQH state at $\nu = \frac 35,\ c=\frac{9}{5}$.  Such a FQH
state is a Fibonacci non-Abelian state.


There are three Fibonacci quasiparticles,
\begin{align}
	\psi_{q_1} = M_2 \ee^{\ii \beta \phi}, \quad \beta = \tfrac{1}{3}\sqrt{3/5}, \quad Q_q = 1/5 \nonumber \\
	\psi_{q_2} = M_1 \ee^{\ii \beta \phi}, \quad \beta = \tfrac{2}{3}\sqrt{3/5}, \quad Q_q = 2/5 \nonumber \\
	\psi_{q_3} = M_0 \ee^{\ii \beta \phi}, \quad \beta = \sqrt{3/5}, \quad Q_q = 3/5 \nonumber \\
\end{align}

The $q_1$ quasiparticle can be calculated from
\begin{align}
	\Psi_{q_1} = \lim_{z_\infty\rightarrow \infty}z_\infty^{h_N}\< \psi_{q_2}^N(z_\infty) \psi_{q_1}(\xi) \psi_e(z_1) \psi_e(z_2)\>
\end{align}
We use $\psi_{q_2}$ as the balancing charge at infinity because $M_2$ has an identity fusion channel with
$M_1$.  Calculating this is more subtle because when we fuse $M_1$ and $M_2$ in the operator expansion, the 
identification of which field belongs to which coordinate is lost.  The series converges to 
$G(z_\infty,\xi,z_1,z_2,z_3) + G(\xi,z_\infty,z_1,z_2,z_3)$, and we choose the term with the correct
scalings for $z_\infty$ and $\xi$.  We obtain
\begin{align}
	G_5(z_\infty,\xi,z_1,z_2,z_3) = \< M_2(z_\infty) M_1(\xi) J_1(z_1) J_1(z_1) J_1(z_1)\> \nonumber \\
	= \frac{1}{6}\mathcal{S}_{z_1,z_2,z_3}\frac{(z_\infty - z_1)(z_\infty - z_2)(\xi-z_3)}{\prod_{i=1}^3(z_\infty-z_i)^{2/3}(\xi-z_i)^{1/3}} G_3(z_1,z_2,z_3)
\end{align}
where $\mathcal{S}_{z_1,z_2,z_3}$ means symmetrize amongst the $z_1,z_2,z_3$ coordinates and $G_3$ is 
the three-point function.  The wavefunction is
\begin{align}
	\Psi_{q_1}(\xi,z_1,z_2,z_3) = \prod_{i<j} (z_i-z_j) \tfrac{1}{3}\sum_{i=1}^3 (\xi - z_i)
\end{align}
The structure here is slightly different than in the previous examples.  Because of the fractional
conformal dimensions, charge $1/3$ is manifested by a single order-1 zero per three electrons.

We can easily obtain the $\Psi_{q_2}$ state by exchanging $z_\infty$ and $\xi$ in $G_5$.
\begin{align}
	\Psi_{q_2}(\xi,z_1,z_2,z_3) = \prod_{i<j} (z_i-z_j) \tfrac{1}{6}\sum_{i\neq j} (\xi - z_i)(\xi - z_j)
\end{align}
This encodes charge $2/3$ by having two zeros for every three electrons.

\section{Hamiltonian realization}

In this section, we construct Hamiltonians such that the FQH states
constructed above are the unique ground states.

First we introduce translation invariant operator $P_{a;L}(z_1,\cdots,z_a)$
that acts on $a$-electron state $|\Psi_a\>$.  The operator is defined as
\begin{align*}
 P_{a;L} =
\begin{cases}
 0, & \text{relative angular momentum of } |\Psi_a\> \geq L\\
 1, & \text{relative angular momentum of } |\Psi_a\> < L\\
\end{cases}
\end{align*}
Then the following Hamiltonian (with a choice of $a_\text{max}$):
\begin{align}
 H_0 = \sum_{a=2}^{a_\text{max}} \sum_{\{i_1,\cdots,i_a\}} P_{a;S_a}(z_{i_1},\cdots,z_{i_a})
\end{align}
will make a FQH state with POZ $S_a$ to have a zero energy.  Such a FQH state
will have a total angular momentum $S_a$ if the state has $a$ electrons.  Also,
any states with less angular momenta will have a positive energy.

Sometimes, $H_0$ already has a unique groundstate.  Other times, different FQH
states may have the same POZ $S_a$; in this case we need an additional
term $H_1$ to select one of the FQH states as the unique ground state. 

For the $h=3/2$ example found earlier, we have the special property that the 
resummation of the operator expansion is the same as the antisymmetrization $\mathcal{A}$ of its singular
terms.  Thus, we can obtain the $n$-body correlation from the 
$(n-2)$-body correlation function using a differential operator and antisymmetrizing
over all particles.  This differential operator for the bare $M_{5,4}$ correlator is
\begin{align}
	{\mathcal{D}}[z_i,z_j] &= \frac{1}{(z_i-z_j)^{3}} 
\\
&\ \ \ \
 + \frac{3/c}{(z_i-z_j)}\left(\sum_{k} \frac{3/2}{(z_k-z_j)^2}
 - \frac{\partial_k}{(z_k-z_j)}\right)
\nonumber 
\end{align}

The correlator of the free boson CFT increases the order of every zero by $\alpha^2$.
To adjust for this, we subtract $\sum_{\ell\neq k}\frac{\alpha^2}{(z_k-z_\ell)}$ 
from the derivative $\partial_k$,
\begin{align}
	{\mathcal{D}}_\alpha[z_i,z_j] = \frac{1}{(z_i-z_j)^{3}} + \frac{3/c}{(z_i-z_j)}\left(\sum_{k\neq i,j} \frac{3/2}{(z_k-z_j)^2}\right. \nonumber \\ 
	\left. - \frac{\partial_k}{(z_k-z_j)} + \sum_{\ell\neq k}\frac{\alpha^2}{(z_k-z_\ell)(z_k-z_j)} \right)
\end{align}

Finally, we can include the Gaussian factor $\ee^{-\sum_i |z_i|^2/4}$ by subtracting $-z^*_i/4$ from
the derivative $\partial_i$,
\begin{align}
	\tilde{\mathcal{D}}_\alpha[z_i,z_j] = \frac{1}{(z_i-z_j)^{3}} + \frac{3/c}{(z_i-z_j)}\left(\sum_{k\neq i,j} \frac{3/2}{(z_k-z_j)^2}\right. \nonumber \\ 
	\left. - \frac{\partial_k+z^*_i/4}{(z_k-z_j)} + \sum_{\ell\neq k}\frac{\alpha^2}{(z_k-z_\ell)(z_k-z_j)} \right)
\end{align}

We can isolate the $(n-2)$-body correlation function from the $n$-body wavefunction
by dividing out a lowest order zero $(z_i-z_j)^{\alpha^2-2h_J}$ and applying a delta 
function $\delta(z_i-z_j)$, which will annihilate all terms with 
higher-order zeros and leave exactly the $(n-2)$-body correlation
function.  Performing these two procedures in succession, we find an operator projecting
to the $n$-body correlation function.  This looks like
\begin{align}
	H_1 = \mathbb{I} - \mathcal{A} \;\tilde{\mathcal{D}}_\alpha[z_i,z_j]\; \delta(z_i-z_j)\frac{1}{(z_i-z_j)^{\alpha^2 - 3}}
\end{align}
We should be careful about the antisymmetrization because there are many different sums
involved.  All the sums within $\tilde{\mathcal{D}}_\alpha$, \ie those over $k$ and $\ell$, should be 
taken with the signs as stated before antisymmetrizing.  When $\tilde{\mathcal{D}}_\alpha$ is applied 
to $(n-2)$-particle wave function, the result is already antisymmetric among $z_k$, with $k\neq i,j$.
The signs differences enter as we change $i$ and $j$.  Antisymmetrization is with respect
to the whole set of coordinates, so this depends on the positions of $i$ and $j$ within the set.
In order to keep the same reference configuration of the other coordinates, we implement this by permuting
$i$ and $j$ to the end of the set and taking the sign of this permutation.  Assuming $i<j$, the sign
of the permutation is $(-1)^{i+j+1}$.  Therefore, we have
\begin{align}
	H_1 = \mathbb{I} - \sum_{i<j} (-1)^{i+j+1} \;\tilde{\mathcal{D}}_\alpha[z_i,z_j]\; \delta(z_i-z_j)\frac{1}{(z_i-z_j)^{\alpha^2 - 3}}
\end{align}
This also encodes the proper normalization via $i<j$, as this will ensure that the most singular terms
occur with coefficent $1$.


Uniqueness of the ground state of this Hamiltonian assumes the wave function is constructed using the iterative
approach of this paper.  In principle, any wavefunction obtained from an arbitrary $(n-2)$-variable
function by applying $\mathcal{D}$ and $\mathcal{A}$ is a ground state of this Hamiltonian.
However, we expect properties of the state not to depend on the exact number of particles in the
system.  Therefore, we argue that if the energetics of an $n$-particle state require it to be 
constructed in this way from the $(n-2)$-particle state, then the energetics of the $(n-2)$-particle
state require it to be constructed in the same way from the $(n-4)$-particle state, and so on.

This particular example is also specific to the $h=3/2$ case because the trick of antisymmetrizing to produce the
correlator from singluar operator expansion terms does not generalize to higher conformal weight.  
However, the $M_{5,4}$ model it arises from
also hosts Fibonacci anyons which are a primary aim of constructing these states, so
we consider it a useful construction.  There is also a state of negative central charge
$c=-21/4$ which could be obtained by the same construction. 

\section{Conclusion}

We have systematically described FQH states that correspond to correlators of $\mathbb{Z}_2$
simple currents in a single-component theory.  Such theories are simple in the regard that they
require only a single Landau level and spin to realize, overcoming energy costs of Landau level
mixing and exchange effects.  They also may require simpler interactions than higher cluster
examples, though complexity also increases with conformal weight.

Our approach is reminiscent of the pattern of zeros (POZ) approach to FQH states in that
we impose only a cluster structure and leading order of zeros.  However, we have gone
further to find that consistency for a theory with central charge imposes a 
rich structure beyond 
the leading order of zero.  Indeed, there are multi-component examples
where the POZ gives the same information and these subleading zeros
fix the central charge and distinguish the theories.  
The subleading zeros also provide useful information in determining 
Hamiltonian realzations of the states.

We find novel 2-cluster states, the $M_{5,4}$ and $M_{6,5}$ states, that support the Fibonacci anyon.
The Fibonacci anyon has previously only been theorized in multi-component or 3-cluster FQH states,
so single-component 2-cluster realizations are novel.  
The $M_{5,4}$ state further has
a candidate four-body parent Hamiltonian, which is encouraging for its realizability.  This eludes
the $M_{6,5}$ 2-cluster state, which has greater complexity owing to its higher conformal weight.
The same topological order may be realized in the 3-cluster $\mathbb{Z}_3$ parafermion state, however, which despite
being 3-cluster has a greatly simplified structure due to its small conformal weight.  The
$M_{5,4}$ state is thus the more novel of the two, and it is in the next tier of wavefunction complexity 
above $\mathbb{Z}_3$ parafermion states.

We would like to thank Shu-Heng Shao for helpful comments. AT was supported by
NSF GRFP grant 2141064.  XGW was supported by NSF grant DMR-2022428 and by the
Simons Collaboration on Ultra-Quantum Matter, which is a grant from the Simons
Foundation (651446).

\appendix

\section{3-cluster $\mathbb{Z}_3$ parafermion state}
\label{Z3}

We provide the 3-cluster realization of the $\mathbb{Z}_3$ parafermion CFT as an illustration of how 
to apply the procedure of the main text beyond 2-cluster states.  Here we have $h = 2/3$.  For definiteness,
we take this state to be at filling $3/5$, so there is a Laughlin factor with $m=5/3$.

For three-cluster states, we must expand in two variables before we can use the Jastrow form of $\tilde f_0$.
We start as in the 2-cluster case with
\begin{align}
	P(z_1,\ldots, z_{3\ell}) = \sum_{k_1,k_2=0}^\infty (z_1-z_2)^{S_2 + k_1}\tilde f_{k_1}(z_2,\ldots,z_{3\ell})
\end{align}
The lowest order of zero in the wavefunction is $S_2 = 1$, consistent with fermionic statistics.  Now we 
expand $f_{k_1}$ in $z_2$ about $z_3$.
\begin{align}
	P(z_1,\ldots, z_{3\ell}) = \sum_{k_1,k_2=0}^\infty (z_1-z_2)^{S_2 + k_1}(z_2-z_3)^{S_3-S_2+k_2}\nonumber \\
	\times \tilde f_{k_1,k_2}(z_2,\ldots,z_{3\ell})
\end{align}
The lowest power of $(z_2-z_3)$ is set so that when combined with $(z_1-z_2)^{S_2}$, we recover
the lowest order of zero $S_3$ for three particles.  For the $\mathbb{Z}_3$ parafermion, this is $S_3=3$,
so the leading power of $(z_2-z_3)$ is $S_3-S_2=2$.

As in the 2-cluster case, we know that $f_{0,0}$ has Jastrow dependence on $z_3$.  Therefore, we factor out the
Laughlin part $\prod_{i<j} (z_i-z_j)^{5/3}$ and work with the associated function $G$,
\begin{align}
	G_{3\ell}(z_1,\ldots, z_{3\ell}) = \sum_{k_1,k_2=0}^\infty (z_1-z_2)^{-2/3 + k_1}(z_2-z_3)^{-4/3+k_2} \nonumber \\
	\times f_{k_1,k_2}(z_3,\ldots,z_{3\ell})
\end{align}
The new leading orders of zero after removing the Laughlin factor are $S_2-m=-2/3$ for two particles and $S_3-3m=-2$ for three particles,
producing the exponents $-2/3$ and $-2+2/3 =-4/3$ in the expansion.  
The function $f$ is defined as
\begin{align}
	&f_{k_1,k_2}(z_3,\ldots,z_{3\ell}) =\prod_{i<j}(z_i-z_j)^{-5/3} \tilde f_{k_1,k_2}(z_3,\ldots,z_{3\ell})\\
	&f_{0,0}(z_3,\ldots,z_{3\ell}) = G(z_4,\ldots,z_{3\ell})
\end{align}

We would like to express $f_{k_1,k_2}$ in terms of operators on $G_{3(\ell - 1)}$.  We will do this in a couple of steps.  Going back
to the expansion just in the first variable $z_1$, we have
\begin{align}
	G_{3\ell}(z_1,\ldots, z_{3\ell}) = \sum_{k_1=0}^\infty (z_1-z_2)^{-2/3 + k_1} f_{k_1}(z_2,\ldots,z_{3\ell})
	\label{expansion1}
\end{align}
We first express $f_{k_1}$ via operators $\phi_{k_1}(z_2)$ acting on an intermediate function $f(z_2,\ldots,z_{3\ell})$.
\begin{align}
	G_{3\ell}(z_1,\ldots, z_{3\ell}) = \sum_{k_1=0}^\infty (z_1-z_2)^{-2/3 + k_1} \phi_{k_1}(z_2) f(z_2,\ldots,z_{3\ell})
	\label{expansion2}
\end{align}
This separates the $k_1$ dependence out from the expansion in the second variable, which we now write as a set of operators
$\phi_{k_2}(z_3)$ acting on $G_{3(\ell-1)}$.
\begin{align}
	f(z_2,\ldots) = \sum_{k_2} (z_2-z_3)^{-4/3+k_2} \phi_{k_2}(z_3) G_{3(\ell-1)}(z_4,\ldots)
	\label{f}
\end{align}

Let us now examine the behavior of $f$ under $\mathcal{L}_0$.  First, we act on Eq.~\ref{expansion2} at the coordinate $z_2$.
Since $G$ has an eigenvalue of $2/3$ under $\mathcal{L}_0$, we have
\begin{align}
	&\left(\mathcal{L}_0(z_2)(z_1-z_2)^{-2/3+k_1}\right)\phi_{k_1}(z_2)f(z_2,\ldots,z_{3\ell})  \nonumber\\
	&+(z_1-z_2)^{-2/3+k_1}[\mathcal{L}_0(z_2),\phi_{k_1}(z_2)] f(z_2,\ldots,z_{3\ell})  \nonumber\\
	&+(z_1-z_2)^{-2/3+k_1} \phi_{k_1}(z_2) \mathcal{L}_0(z_2)f(z_2,\ldots,z_{3\ell}) \nonumber\\
	&= 2/3(z_1-z_2)^{-2/3+k_1} \phi_{k_1}(z_2) f(z_2,\ldots,z_{3\ell})\\
	&2/3 - k_1 - 2/3 + k_1 + h_{z_2} = 2/3
\end{align}
This shows that the eigenvalue $h_{z_2}$ of $f$ under $\mathcal{L}_0$ under $z_2$ is $2/3$.

We can do the same with $z_3$ by applying $\mathcal{L}_0(z_3)$ to Eq.~\ref{f}.
\begin{align}
	&h_{z_3}f(z_2,\ldots) \nonumber \\
	&\;\;= \mathcal{L}_0(z_3)\sum_{k_2} (z_2-z_3)^{-4/3+k_2} \phi_{k_2}(z_3) G_{3(\ell-1)}(z_4,\ldots)
\end{align}
Distributing $\mathcal{L}_0$ in a similar way, we obtain
\begin{align}
	&h_{z_3} = 4/3 - k_2 - 2/3 + k_2 = 2/3
\end{align}
It is straightforward to see that $h_{z_i}$ for $i>3$ inherit the value $2/3$ from $G$.  Therefore $\mathcal{L}_0(z_i)f(z_2,\ldots,z_{3\ell})=2/3$
for all coordinates $z_i$.

The operators $\phi_{k_1}$ and $\phi_{k_2}$ are once again in the basis of Virasoro operators at level $k_1$ and $k_2$, respectively.  
\begin{align}
	\phi_{k_i} = \sum_{n_1+\ldots+n_j=k_i} \beta_i^{\{n_1,\ldots,n_j\}}\prod_{p=1}^j \mathcal{L}_{-n_p}
\end{align}
The coefficients $\beta$ may be different for $\phi_{k_1}$ and $\phi_{k_2}$, so we have labeled them with $i$ for $i=1,2$.
We fix them by applying an $L_+$ operator and requring this to annihilate the state.  First, we apply $L_+$ at the coordinate
$z_2$ in Eq.~\ref{expansion2}.  As in the main text, we can split it into $\mathcal{L}_{n}(z_2;z_1)$ acting differentially
on the first coordinate and $\bar{\mathcal{L}}_{n}(z_2;z_3,\ldots)$ which commute through the $\phi_k$ via Virasoro commutators.
As another assumption, we will take $L_+$ to also annihilate $f(z_2,\ldots,z_{3\ell})$.  We obtain the equation
\begin{align}
	(2/3(n+1)-2/3+k_1)\phi_{k_1} = [\bar{\mathcal{L}_n},\phi_{k_1+n}]
	\label{ope_constraint1}
\end{align}
This can be used to iteratively fix the coefficients $\beta_1$ at level $k+n$ from the coefficients at level $k$.  We find that the resulting
$\phi_k$ match the OPE for two primary fields of weight $h_{1,2}=2/3$ that fuse to another primary field of weight $h_3=2/3$.  The coefficients
of the OPE would be fixed by an equation similar to Eq~\ref{ope_constraint1}, where the left hand side of equation~\ref{ope_constraint1} would 
have $(h_1(n+1)-h_1-h_2+h_3+k)\phi_k$.

We can similarly fix $\phi_{k_2}$ by acting with $L_+$ via $\mathcal{L}_n(z_3;z_2)+\bar{\mathcal{L}}_n(z_3;z_4,\ldots)$ on Eq.~\ref{f}.
The condition is
\begin{align}
	(2/3(n+1)-4/3+k_1)\phi_{k_1} = [\bar{\mathcal{L}_n},\phi_{k_1+n}]
	\label{ope_constraint2}
\end{align}
This matches what we expect from CFT for an OPE of two primary fields of conformal weight $2/3$ that fuse to the vaccum.

Now it remains to fix the central charge.  We start with the three-particle function,

\begin{align}
	G_3(z_1,z_2,z_3) =\sum_{\{n_k\}} \frac{\beta^{\{n_k\}}}{(z_1-z_2)^{2/3-\sum_k n_k}}\nonumber\\
	\left(\prod_k\mathcal{L}_{n_k}(z_2;z_3) \right)\frac{1}{(z_2-z_3)^{4/3}}
\end{align}
	
This operator expansion matches the expansion of 
\begin{align}
	G_3(z_1,z_2,z_3) = \frac{1}{(z_1-z_2)^{2/3}(z_1-z_3)^{2/3}(z_2-z_3)^{2/3}}.
\end{align}
Interestingly, this turns out not to depend on $c$.  Computing terms in the expansion, factors of $c$
empirically drop out upon simplifying.  We must next find the five-particle function $f$ in order to fix $c$.
\begin{align}
	\label{G5}
	f&(z_1,\ldots,z_5) \nonumber \\
	&= \sum_{\{n_k\}} \frac{\beta^{\{n_k\}}}{(z_1-z_2)^{4/3-\sum_k n_k}}\left(\prod_k\mathcal{L}_{n_k}(z_2;z_3,z_4,z_5) \right) \nonumber \\
	&\;\;\;\;\;\;\frac{1}{(z_3-z_4)^{2/3}(z_3-z_5)^{2/3}(z_4-z_5)^{2/3}}
\end{align}

Determining the highest order poles in $f$ is slightly tricky.  Looking back at the series in Eq.~\ref{expansion1}, it is clear that
while most of the coordinates should have a straightforward exchange symmetry, the first coordinate in $f$ is distinguished via the expansion.
The coordinates which have exchange symmetry should have highest order $2/3$ poles from the pattern of zeros.  We see that $z_1$ and $z_2$
have a $4/3$-order pole in the expansion, and this order should also appear between $z_1$ and each other coordinate to enforce associativity.
Therefore we have the overall factor of highest-order poles,
\begin{align}
	\tilde{G}_5(z_1,\ldots,z_5) = \frac{1}{\prod_{i=2}^5 (z_1-z_i)^{4/3}\prod_{j=2,i<j}^5 (z_i-z_j)^{2/3}}
\end{align}
We must find the symmetric polynomial $P_5(z_1,\ldots,z_5)$ such that the scaling of each coordinate in $f(z_1,\ldots,z_5)=\tilde{G}(z_1,\ldots,z_5)P_5(z_1,\ldots,z_5)$
is $-2h=-4/3$.  It turns out that there is only one possibility,
\begin{align}
	P_5(z_1,\ldots, z_5) = \mathcal{S}_{z_1} \left( (z_1-z_2)^2 (z_1-z_3)^2 (z_4-z_5)^2\right)
\end{align}
Here $\mathcal{S}_{z_1}$ indicates that all coordinates except the distinguished coordinate $z_1$ are symmetrized.

There is one subtlety.  Calculations show that the operator expansion in Eq.~\ref{G5} has a symmetry between
$z_1$ and $z_2$.  In other words, in writing the expansion we have distinguished $z_1$ or $z_2$ as carrying
the $4/3$-order poles, but the series produces the same result whether we use $\phi_{k_2}(z_1)$ or $\phi_{k_2}(z_2)$.
This can be understood from the CFT perspective that once fields $\phi_1(z_1)$ and $\phi_2(z_2)$ fuse to the 
vacuum, we lose information about which coordinate belongs to which field.  Empirically, the series in Eq.~\ref{G5} matches a symmetrization
of the function $\tilde{G}_5P_5$ between $z_1$ and $z_2$. Thus we define
\begin{align}
	\tilde{G'}_5(z_1,\ldots,z_5) = \frac{1}{\prod_{i=2}^5 (z_2-z_i)^{4/3}\prod_{j\neq 2,i<j} (z_i-z_j)^{2/3}}\\
	P'_5(z_1,\ldots, z_5) = \mathcal{S}_{z_2} \left( (z_1-z_2)^2 (z_2-z_3)^2 (z_4-z_5)^2\right)
\end{align}
The result for $f$ is then
\begin{align}
	f = \tilde G_5 P_5 + \tilde G'_5 P'_5
\end{align}
Matching this with the expansion, we find a result of $c=4/5$, which agrees with the $\mathbb{Z}_3$ parafermion CFT.

We can now go on to compute the six-particle function $G_6$ for completeness. Here we use only the term in $G_5$ corresponding
to the correct pairing of fields and coordinates, $\tilde G_5 P_5 $.

\begin{align}
	\label{G5}
	&G_6(z_1,\ldots,z_6) =\nonumber \\
	&\sum_{\{n_k\}} \frac{\beta^{\{n_k\}}}{(z_1-z_2)^{2/3-\sum_k n_k}}\left(\prod_k\mathcal{L}_{n_k}(z_2) \right) \tilde G_5 P_5
\end{align}
The prefactor is a straightforward product,
\begin{align}
	\tilde G_6(z_1,\ldots,z_6) = \frac{1}{\prod_{i<j} (z_i-z_j)^{2/3}}
\end{align}
There is again only one possibility for the numerator polynomial $P_6$, which must have a highest
power of 2 on every coordinate.  This is
\begin{align}
	P_6(z_1,\ldots, z_6) = \mathcal{S}\left((z_1-z_2)^2(z_3-z_4)^2(z_5-z_6)^2\right)
\end{align}
The product $G_5 = \tilde{G}_6 P_6$ is exactly the 3-cluster $\mathbb{Z}_3$ parafermion state, which is what we expected.

\section{ 4-point correlation of $\Z_2$ simple current operators in the $M_{5,4}$ model}

The 4-point function for a minimal model can be calculated using a null vector equation.  Here we perform this calculation for
the $M_{5,4}$ model.  The approach is as follows:

A null vector is a zero eigenvector of the Gram matrix, the matrix of overlaps of the descendant 
fields at a certain level.  By definition, a null vector has zero norm and is orthogonal to the rest of the descendant fields at that level.
Descendant fields at different levels are also orthogonal to each other, so the null vector is orthogonal
to the entire Verma module.  Such a field, inserted into any correlator, then produces zero.  Using the differential operator
expressions for the Virasoro operators inserted into the correlation function, we can then write down a homogeneous differential equation
for any correlator of primary fields.  For a 4-point function, we can also use a conformal transformation to fix three of the 
four coordinates, so this differential equation becomes an ODE in the final coordinate, which can then be solved to find the 4-point function.

To begin, we must find a null vector for the $\Z_2$ simple current $J$ of the $M_{5,4}$ model.  This first null vector for this field occurs at level 3, so we need
the zero eigenvector of the Gram matrix of the level-3 basis $\{ L_{-1}L_{-1}L_{-1}|J\>, L_{-2}L_{-1}|J\>, L_{-3}|J\>\} \equiv U$.  Using the commutators of the
Virasoro algebra and plugging in values of $c=7/10$ and $h=3/2$, we find
\begin{align}
	U^\dagger U =
	\begin{pmatrix}
		 360 &    99    & 36  \\
		 99  &  621/20  & 15  \\
		  36  &  15   &  52/5 \\
	\end{pmatrix}
\end{align}
This has a zero eigenvector $\eta = (1,-5,15/4)$.  The descendant field that is orthogonal to the rest of the module is
\begin{align}
	\eta_1 L_{-1}L_{-1}L_{-1} J + \eta_2 L_{-2} L_{-1} J + \eta_3 L_{-3} J 
\end{align}
The Virasoro operators $L_{-n}$ correspond to differential operators $\mathcal{L}_{-n}$,
\begin{align}
	\mathcal{L}_{-n} = \sum_i \left( \frac{(n-1)h_i}{(z_i-z)^n} - \frac{\partial_i}{(z_i-z)^{n-1}}\right)
\end{align}
where $z$ is the coordinate of the field $J$ the Virasoro operators are acting on, and $z_i$ are the coordinates of
all other primary fields in the correlator.

For the four-point function $\< J(z_1)J(z_2)J(z_3)J(z_4)\>$, we will take the Virasoro generators to act on $J(z_3)$
 and $i$ runs over $1,2,4$.  The next step is to apply a conformal map to fix $z_1,z_2,z_4$ to $0,1,\infty$.  The 
map is
\begin{align}
	w \mapsto \frac{(w-z_1)(z_2-z_4)}{(w-z_4)(z_2-z_1)}
\end{align}
We define the image of $z_3$ under this map, $\frac{(z_3-z_1)(z_2-z_4)}{(z_2-z_1)(z_3-z_4)} \equiv z$.
The form of the four-point function is fixed by conformal symmetry to a prefactor encoding the scaling dimensions times a function 
of this cross-ratio,
\begin{align}
	\< J(z_1)J(z_2)J(z_3)J(z_4)\> = \frac{1}{(z_1-z_2)^{2h}(z_3-z_4)^{2h}} G(z)
\end{align}
Under the conformal map, this becomes formally infinite; however, we can still derive a differential equation for the 
undetermined function $G(z)$, which conveniently maps to itself.  We start by studying the action of partial derivatives
on the original form of the four-point function, $F(z_1,z_2,z_3,z_4) \equiv \< J(z_1)J(z_2)J(z_3)J(z_4)\> =\frac{1}{(z_1-z_2)^{2h}(z_3-z_4)^{2h}} G(z)$.
\begin{align}
	\partial_1 F = \left( \frac{-2h}{(z_1-z_2)} + \frac{(z_3-z_2)(z_2-z_4)}{(z_2-z_1)^2(z_3-z_4)}\partial_z \right) F\\
	\partial_2 F = \left( \frac{2h}{(z_1-z_2)} + \frac{(z_4-z_1)(z_3-z_1)}{(z_2-z_1)^2(z_3-z_4)}\partial_z \right) F \\
	\partial_3 F = \left( \frac{-2h}{(z_3-z_4)} + \frac{(z_1-z_4)(z_2-z_4)}{(z_2-z_1)(z_3-z_4)^2}\partial_z\right) F \\
	\partial_4 F = \left( \frac{2h}{(z_3-z_4)} + \frac{(z_3-z_1)(z_2-z_3)}{(z_2-z_1)(z_3-z_4)^2}\partial_z\right) F 
\end{align}
In this form, with the $\partial_z$ derivatives acting only on $G(z)$, we can straightforwardly apply the conformal map by plugging in values for
$z_1,z_2,z_3,z_4$.  Thus we can automatically find
\begin{align}
	\mathcal{L}_{-3} = &\frac{-2h}{z^3}  - \frac{1}{z^2} (2 h  + (z - 1)  \partial_z ) \\
	&+ \frac{2 h}{(1 - z)^3}  -  \frac{1}{(1 - z)^2} (-2 h  - z  \partial_z)
\end{align}
For the other two descendants $\mathcal{L}_{-2}\mathcal{L}_{-1}$ and $\mathcal{L}_{-1}^3$, we have to be careful
to move all the derivatives onto $G(z)$ before mapping because derivatives acting on $z_1,z_2,z_4$ do not commute with the conformal map.  In both of these cases,
the operators to the right are $\mathcal{L}_{-1} = \partial_3$, so the relevant derivative is $\partial_i$ acting on the 
prefactor of in $\partial_3$.  This derivative factor goes to zero under the conformal map for $\partial_3$ and $\partial_4$, but contributes an added $1$ and $-1$
for $\partial_1$ and $\partial_2$, respectively.  There is no net effect for $\mathcal{L}_{-1}^3$, but for $\mathcal{L}_{-2}\mathcal{L}_{-1}$
the shift becomes apparent.  The result is
\begin{align}
	\mathcal{L}_{-1}^3 = \partial_z^3 \\
	\mathcal{L}_{-2}\mathcal{L}_{-1} = &\left(\frac{h}{z^2} + \frac{1}{z} ((2 h + 1)  + (z - 1) \partial_z)\right.\nonumber\\ 
	&\left.+ \frac{h}{(1 - z)^2} + \frac{1}{(z - 1)} (-(2 h + 1) - z \partial_z)\right)\partial_z
\end{align}
We can now piece together the ODE for $G(z)$,
\begin{align}
	&\left[ \eta_1 \partial_z^3 + \eta_2\left(\frac{h}{z^2} + \frac{1}{z} ((2 h + 1)  + (z - 1) \partial_z)\right.\right.\nonumber\\ 
	&\;\left.+ \frac{h}{(1 - z)^2} + \frac{1}{(z - 1)} (-(2 h + 1) - z \partial_z)\right)\partial_z \nonumber\\
	&\;+ \eta_3\left(\frac{-2h}{z^3}  - \frac{1}{z^2} (2 h  + (z - 1)  \partial_z )\right. \nonumber\\
	&\;\left.\left.+ \frac{2 h}{(1 - z)^3}  -  \frac{1}{(1 - z)^2} (-2 h  - z  \partial_z)\right)\right] G(z) = 0
\end{align}
This equation can be solved efficiently by noticing that $G(z)$ must be antisymmetric under $z \rightarrow (1-z)$ (corresponds to swapping $z_1$ and $z_2$)
and $z\rightarrow 1/z$ picking up an additional factor $1/z^{2h}$ (corresponds to swapping $z_1$ and $z_4$).  The $z\rightarrow 1/z$ symmetry constrains
the powers allowed to appear in the solution quite heavily, since to have $G(z) = -\frac{1}{z^{2h}}G(1/z)$ with $G(z)$ decaying to infinity, we can only have
powers of $z$ between $0$ and $-2h$.  These powers must also be integer for the correlation function to be single-valued.  For our case, $h=3/2$, so we can
quickly check all functions $z^a(1-z)^b$ for $a,b=0,-1,-2,-3$ antisymmetrized under the above transformations.  We find a solution using only two of them,
$\mathcal{A}(z^{-3})$ and $\mathcal{A}(z^{-1})$ where $\mathcal{A}$ denotes the antisymmetrization,
\begin{align}
	G(z) = \mathcal{A}\left(\frac{1}{z^3}\right) + \frac{30}{7}\mathcal{A}\left(\frac{1}{z}\right)
\end{align}

Including the prefactor and writing out the antisymmetrization, this becomes
\begin{align}
	\< J(z_1)J(z_2)J(z_3)J(z_4)\> = \frac{2}{(z_1-z_2)^3(z_3-z_4)^3} \nonumber \\
	- \frac{2}{(z_4-z_2)^3(z_3-z_1)^3} + \frac{2}{(z_4-z_1)^3(z_3-z_2)^3} \nonumber \\
	+ \frac{30}{7} \frac{3}{\prod_{i<j} (z_i-z_j)}
\end{align}

One can verify that this function matches the operator expansion result
\begin{align}
	\< &J(z_1)J(z_2)J(z_3)J(z_4)\> = \mathcal{A}\left( \left[\frac{1}{(z_1-z_2)^3}\right.\right.\nonumber\\ 
	&\left.\left.+ \frac{2h/c}{(z_1-z_2)}\sum_{i=3}^4\left(\frac{h}{(z_i-z_2)^2} - \frac{\partial_i}{(z_i-z_2)}\right)\right]\frac{1}{(z_3-z_4)^3} \right)
\end{align}
where $2h/c = 30/7$.

\section{Multi-component states}
We have highlighted single component (spinless) candidates for states hosting Fibonacci anyons
in the main text; however there are also multi-component options which we detail here.
\subsection{$su(2)_3$ realization of a $\Z_2$ simple current algebra}

Actually there
are two such states, $su(2)_3$-state and $sp(6)_1$-state, with different chiral
central charge $c=\frac {14}{5}$ and $ \frac{26}{5}$.  Here the chiral central
charge is the number of right-moving edge modes minus that of left-moving edge
modes in the FQH state.  

A $\Z_p$ simple current algebra
\[
J_r \times J_s \;\sim\; J_{\,r+s \,\text{mod}\,p},\qquad J_0 \equiv \mathbf{1},
\]
is not only characterized by the conformal weights of the simple currents,
but also by the central charge \cite{LWW1024}.  
Let us first consider a $\Z_2$ simple current algebra with
\[
h(J_1)=\tfrac{3}{4},\qquad c=\tfrac{9}{5}.
\]

This algebra arises in the $su(2)_3$ Kac-Moody CFT, which has four primary fields:
\\[2mm]
\centerline{
\begin{tabular}{|c|c|c|c|c|}
\hline
primary fields & $\onebb$ & $J_1$ & $M_0$ & $M_1$ \\ 
\hline
$h$ & $0$ & $\tfrac{3}{4}$ & $\tfrac{2}{5}$ & $\tfrac{3}{20}$ \\ 
\hline
$d$ & $1$ & $1$ & $\tfrac{1+\sqrt{5}}{2}$ & $\tfrac{1+\sqrt{5}}{2}$ \\ 
\hline
\end{tabular}
}
\vskip 2mm

The $S$-matrix is
\begin{align}
S=\begin{pmatrix}
1 & 1 & \tfrac{1+\sqrt{5}}{2} & \tfrac{1+\sqrt{5}}{2} \\ 
1 & -1 & \tfrac{1+\sqrt{5}}{2} & -\tfrac{1+\sqrt{5}}{2} \\ 
\tfrac{1+\sqrt{5}}{2} & \tfrac{1+\sqrt{5}}{2} & -1 & -1 \\ 
\tfrac{1+\sqrt{5}}{2} & -\tfrac{1+\sqrt{5}}{2} & -1 & 1 
\end{pmatrix}.
\end{align}

The fusion rules are
\\[2mm]
\centerline{
\begin{tabular}{ |c||c|c|c|c|}
 \hline 
$\otimes$  & $\onebb$  & $J_1$  & $M_0$  & $M_1$ \\ 
\hline\hline 
$\onebb$  & $\onebb$  & $J_1$  & $M_0$  & $M_1$  \\ 
\hline 
$J_1$     & $J_1$     & $\onebb$ & $M_1$  & $M_0$  \\ 
\hline 
$M_0$     & $M_0$     & $M_1$    & $\onebb \oplus M_0$ & $J_1 \oplus M_1$ \\ 
\hline 
$M_1$     & $M_1$     & $M_0$    & $J_1 \oplus M_1$ & $\onebb \oplus M_0$ \\ 
\hline 
\end{tabular}
}
\vskip 2mm

Using the simple current $J_1$, the electron operator is
\begin{align}
 \psi_e = J_1 \, \ee^{\ii \alpha \phi}, \qquad \alpha=\sqrt{\tfrac{3}{2}+1},
\end{align}
which produces a FQH state at 
\[
\nu=\tfrac{2}{5},\qquad c=\tfrac{14}{5}.
\]
We refer to this as the $su(2)_3$ state, a Fibonacci non-Abelian state.

The minimally charged quasiparticle is
\begin{align}
 \psi_q = M_1 \, \ee^{\ii \beta \phi}, \qquad \beta=\tfrac{1}{2\alpha}=\tfrac{1}{\sqrt{10}},
\end{align}
with electric charge
\begin{align}
 Q_q = \tfrac{\beta}{\alpha} = \tfrac{1}{5}
\end{align}
and quantum dimension $d= \tfrac{1+\sqrt{5}}{2}$. The anyon has Fibonacci
non-Abelian statistics.

\subsection{$sp(6)_1$ realization of a $\Z_2$ simple current algebra}

Another $\Z_2$ simple current algebra is specified by
\[
h(J_1)=\tfrac{3}{4},\qquad c=\tfrac{21}{5}.
\]
It appears in the $sp(6)_1$ Kac-Moody CFT, which again has four primary fields:
\\[2mm]
\centerline{
\begin{tabular}{|c|c|c|c|c|}
\hline
primary fields & $\onebb$ & $J_1$ & $M_0$ & $M_1$ \\ 
\hline
$h$ & $0$ & $\tfrac{3}{4}$ & $\tfrac{3}{5}$ & $\tfrac{17}{20}$ \\ 
\hline
$d$ & $1$ & $1$ & $\tfrac{1+\sqrt{5}}{2}$ & $\tfrac{1+\sqrt{5}}{2}$ \\ 
\hline
\end{tabular}
}
\vskip 2mm

The $S$-matrix and fusion rules are identical to those in the $su(2)_3$ case.  

The electron operator is
\begin{align}
 \psi_e = J_1 \, \ee^{\ii \alpha \phi}, \qquad \alpha=\sqrt{\tfrac{3}{2}+1},
\end{align}
which yields an FQH state at
\[
\nu=\tfrac{2}{5},\qquad c=\tfrac{26}{5}.
\]
We call this the $sp(6)_1$ state, which again has Fibonacci non-Abelian
statistics.

The minimal-charge quasiparticle is
\begin{align}
 \psi_q = M_1 \, \ee^{\ii \beta \phi}, \qquad \beta=\tfrac{1}{\sqrt{10}},
\end{align}
carrying charge
\[
Q_q=\tfrac{1}{5}
\]
and quantum dimension $d= \tfrac{1+\sqrt{5}}{2}$. It is an anyon with Fibonacci
non-Abelian statistics.

\subsection{$su(2)_5$ realization of a $\Z_2$ simple current algebra}

Finally, consider the $su(2)_5$ Kac-Moody CFT, with
\[
h(J_1)=\tfrac{5}{4},\qquad c=\tfrac{15}{7}.
\]
This theory contains six primary fields:
\\[2mm]
\centerline{
\begin{tabular}{|c|c|c|c|c|c|c|}
\hline
primary fields & $\onebb$ & $J_1$ & $M_0$ & $M_1$ & $M_2$ & $M_3$ \\ 
\hline
$h$ & $0$ & $\tfrac{5}{4}$ & $\tfrac{6}{7}$ & $\tfrac{3}{28}$ & $\tfrac{2}{7}$ & $\tfrac{15}{28}$ \\ 
\hline
$d$ & $1$ & $1$ & $-c_{7}^{3}$ & $-c_{7}^{3}$ & $\xi_{7}^{3}$ & $\xi_{7}^{3}$ \\ 
\hline
\end{tabular}
}
\vskip 2mm
where
\begin{align}
&
\zeta^m_n=\mathrm{e}^{2\pi \mathrm{i} m/n}, \ \ \
\xi^{m}_n = (\zeta^m_{2n}-\zeta^{-m}_{2n})/(\zeta_{2n}-\zeta_{2n}^{-1})\nonumber \\
&c^m_n = \zeta^m_n+\zeta^{-m}_n,\ \ \
\end{align}

The corresponding $S$-matrix is given by
\[
S=
\begin{pmatrix}
1,
& 1,
& -c_{7}^{3},
& -c_{7}^{3},
& \xi_{7}^{3},
& \xi_{7}^{3} \\ 
1,
& -1,
& -c_{7}^{3},
& c_{7}^{3},
& \xi_{7}^{3},
& -\xi_{7}^{3} \\ 
-c_{7}^{3},
& -c_{7}^{3},
& -\xi_{7}^{3},
& -\xi_{7}^{3},
& 1,
& 1 \\ 
-c_{7}^{3},
& c_{7}^{3},
& -\xi_{7}^{3},
& \xi_{7}^{3},
& 1,
& -1 \\ 
\xi_{7}^{3},
& \xi_{7}^{3},
& 1,
& 1,
& c_{7}^{3},
& c_{7}^{3} \\ 
\xi_{7}^{3},
& -\xi_{7}^{3},
& 1,
& -1,
& c_{7}^{3},
& -c_{7}^{3} \\ 
\end{pmatrix}
\]
and the fusion rule is given by\\[2mm]
\centerline{\scriptsize
\begin{tabular}{ |c||c|c|c|c|c|c|}
 \hline 
$\otimes$  & $\onebb$  & $J_1$  & $M_0$  & $M_1$  & $M_2$  & $M_3$ \\ 
\hline 
 \hline 
$\onebb$  & $ \onebb$  & $ J_1$  & $ M_0$  & $ M_1$  & $ M_2$  & $ M_3$  \\ 
 \hline 
$J_1$  & $ J_1$  & $ \onebb$  & $ M_1$  & $ M_0$  & $ M_3$  & $ M_2$  \\ 
 \hline 
$M_0$  & $ M_0$  & $ M_1$  & $ \onebb \oplus M_2$  & $ J_1 \oplus M_3$  & $ M_0 \oplus M_2$  & $ M_1 \oplus M_3$  \\ 
 \hline 
$M_1$  & $ M_1$  & $ M_0$  & $ J_1 \oplus M_3$  & $ \onebb \oplus M_2$  & $ M_1 \oplus M_3$  & $ M_0 \oplus M_2$  \\ 
 \hline 
$M_2$  & $ M_2$  & $ M_3$  & $ M_0 \oplus M_2$  & $ M_1 \oplus M_3$  & $ \onebb \oplus M_0 \oplus M_2$  & $ J_1 \oplus M_1 \oplus M_3$  \\ 
 \hline 
$M_3$  & $ M_3$  & $ M_2$  & $ M_1 \oplus M_3$  & $ M_0 \oplus M_2$  & $ J_1 \oplus M_1 \oplus M_3$  & $ \onebb \oplus M_0 \oplus M_2$  \\ 
 \hline 
\end{tabular}
}
\vskip 2mm
The electron operator
\begin{align}
 \psi_e = J_1 \, \ee^{\ii \alpha \phi}, \qquad \alpha=\sqrt{\tfrac{5}{2}+1},
\end{align}
generates an FQH state at
\[
\nu=\tfrac{2}{7},\qquad c=\tfrac{22}{7}.
\]
We call this the $su(2)_5$ state, which realizes a new class of non-Abelian statistics beyond Ising and Fibonacci.

The minimal-charge quasiparticle is non-Abelian:
\begin{align}
 \psi_q = M_1 \, \ee^{\ii \beta \phi}, \qquad \beta=\tfrac{1}{2\alpha}=\tfrac{1}{\sqrt{14}},
\end{align}
which carries a charge
\[
Q_q=\tfrac{1}{7}.
\]

\

\

\bibliography{all,publst}

\end{document}